\documentclass[journal]{IEEEtran}
\makeatother
\usepackage[ruled,vlined]{algorithm2e}
\makeatletter
\renewcommand{\@algocf@capt@plain}{above}
\makeatother
\usepackage{fonttable}
\usepackage{amsfonts}
\usepackage{graphicx} 
\usepackage{url}
\usepackage[caption=false,font=footnotesize]{subfig}
\usepackage{amsbsy}
\usepackage{amsmath, amssymb, bm}
\usepackage{graphicx,xparse}
\usepackage{mathtools}
\usepackage{placeins}
\usepackage{pdfpages}
\usepackage{mathptmx}
\usepackage{mathtools}
\usepackage{mathrsfs} 
\usepackage{multirow}
\usepackage{array}
\usepackage[utf8]{inputenc}
\usepackage{acronym}
\usepackage{mathptmx}
\usepackage{enumitem}
\setlist[itemize]{noitemsep, topsep=0pt}
\setlist[enumerate]{noitemsep, topsep=0pt}
\usepackage{longtable}
\usepackage[hidelinks]{hyperref}
\usepackage{xcolor}
\hypersetup{
    colorlinks,
    linkcolor={black!50!black},
    citecolor={black!90!black},
    urlcolor={black!80!black}
}
\usepackage{xcolor}
\usepackage{cite}
\usepackage{comment}
\usepackage[acronym]{glossaries}

\newacronym{6g}{6G}{sixth generation}
\newacronym{ris}{RIS}{reconfigurable intelligent surfaces}
\newacronym{dnn}{DNN}{Deep Neural Network}
\newacronym{ann}{ANN}{Artificial Neural Network}
\newacronym{los}{LoS}{line-of-sight}
\newacronym{so}{SOs}{scattering objects}
\newacronym{bilstm}{biLSTM}{bidirectional long-short term memory}
\newacronym{lstm}{LSTM}{long-short term memory}
\newacronym{ue}{UE}{user equipment}
\newacronym{rmse}{RMSE}{root mean squared error}
\newacronym{bs}{BS}{base station}
\newacronym{snr}{SNR}{signal-to-noise ratio}
\newacronym{rse}{RSE}{rich-scattering environment}
\newacronym{rses}{RSEs}{rich-scattering environments}
\newacronym{dl}{DL}{deep learning}
\newacronym{ml}{ML}{machine learning}
\newacronym{rnn}{RNN}{recurrent neural network}
\newacronym{mimo}{MIMO}{multiple-input-multiple-output}
\newacronym{siso}{SISO}{single-input-single-output}
\newacronym{crlb}{CRLB}{Cramér-Rao lower bound}
\newacronym{bcrlb}{BCRLB}{Bayesian Cramér-Rao lower bound}
\newacronym{ofdm}{OFDM}{frequency division
multiplexing}
\newacronym{rss}{RSS}{received signal strength}
\newacronym{miso}{MISO}{multiple-input-single-output}
\newacronym{bo}{BO}{Bayesian Optimization}
\newacronym{bisen}{Bi-SEN}{Scattering Estimation Network}
\newacronym{biaruln}{Bi-ARULN}{Adaptive RIS-Assisted User Localization Network}
\newacronym{cdf}{CDF}{cumulative distribution function}
\newacronym{aoa}{AoA}{angle-of-arrival}
\newacronym{toa}{ToA}{time-of-arrival}
\newacronym{fim}{FIM}{Fisher Information Matrix}
\newacronym{rf}{RF}{Radio Frequency}

\title{Integrated Sensing, User Location and Orientation Estimation in RIS-Assisted Dynamic Rich Scattering Environment}
\author{Anum~Umer, Ivo~M\"{u}\"{u}rsepp,~\IEEEmembership{Member,~IEEE}, Muhammad~Mahtab~Alam,~\IEEEmembership{Senior~Member,~IEEE}
\thanks{A. Umer, I. M\"{u}\"{u}rsepp, and M. M. Alam are with Thomas Johann Seebeck Department of Electronics, Tallinn University of Technology, 19086 Tallinn, Estonia. (e-mail: anum.umer@taltech.ee, ivo.muursepp@taltech.ee, muhammad.alam@taltech.ee). This project has received funding from the European Union’s Horizon Europe Research Program under grant agreement No. 101058505 - 5G-TIMBER, and from the Estonian Education and Youth Board ÕÜF11 \"AIoT*5G - Artificial intelligence, edge computing and IoT solutions in distributed systems.}}

\begin{document}
\maketitle

%

\begin{abstract} 
This paper investigates an uplink \acrfull{ue} location and orientation estimation problem in an indoor \acrfull{rse} for a \acrfull{mimo} narrowband \acrfull{ris}-assisted communication system. The localization problem in \acrshort{rse} is challenging as the uplink pilot signal undergoes multiple interactions with the \acrshort{ris} and dynamic \acrfull{so}. This paper proposes an approach where \acrfull{bs} adaptively senses the environment with the help of \acrshort{ris}. Based on this sensing,  it sequentially designs \acrshort{ris} configuration, \acrshort{bs} beamforming and \acrshort{ue} beamforming vectors, using the sequence of pilot transmissions from the \acrshort{ue} to the \acrshort{bs}, with an objective of progressively focusing them onto the \acrshort{ue}. Towards this end, we train a \acrfull{bilstm} network based controller to capture the temporal dependencies between measurements to first adaptively sense the \acrshort{rse} and then design \acrshort{ris}, \acrshort{bs} and \acrshort{ue} beamforming vectors to localize the \acrshort{ue}. We evaluate the proposed approach under various \acrshort{rse} conditions such as various distributed \acrshort{ris} installations, varying number of randomly moving \acrshort{so} and sensing \acrshort{ris} elements. Simulation results illustrate that it effectively enables adaptive sensing to achieve low localization error with robustness in various \acrshort{rses}.
\end{abstract}


\section{Introduction}
Achieving reliable localization in \acrfull{rses} dominated by dense multipath and rapid changes has become increasingly important as many modern applications depend on precise and resilient positioning \cite{9847080,bourdoux20206g}. Conventional approaches often fail in such settings, especially indoors, where infrastructure constraints and rich scattering further complicate signal behavior. The need is amplified in emerging 6G networks that span microwave through terahertz and optical bands, where indoor environments exhibit strong scattering \cite{you2021towards}.  The \acrfull{ue} localization becomes challenging in narrowband indoor environments with rich scattering multipath propagation, where the \acrshort{ue} pilot signal is reflected by surrounding objects and reaches the \acrfull{bs} through multiple paths, after several reflections. As a result, these overlapping propagation components with different amplitudes and phases are hard to distinguish from one another.

This paper explores the role of \acrfull{ris} as an affordable substitute for large antenna arrays for localizing the \acrshort{ue} in \acrshort{rses}. The \acrshort{ris} use numerous passive low power elements to tune reflected wave phases and regulate the propagation environment \cite{9140329,PhysRevApplied.11.044024,li2017electromagnetic}. These characteristics enable them to act as intelligent anchors or enhance existing systems to mitigate interference, and directional issues to support the localization requirements for autonomous platforms, robotics, immersive applications, smart manufacturing, and traffic management \cite{9847080,9548046,bourdoux20206g,10044963,10858311}.
It is viewed as a key enabler for 6G localization because it can maintain a dependable reflected link when the direct path is weak or obstructed and can introduce additional anchor points without the need for extra \acrfull{rf} chains \cite{bourdoux20206g, 10858311}. Under dynamic \acrshort{rse} conditions, repeated interactions with objects and RIS elements create composite reflections with diverse \acrfull{aoa} and polarizations \cite{9856592,10453467,10133065}, leading to nonlinear behavior far from free or quasi free space models \cite{10858311}. In these settings, RIS supply controlled perturbations, whereas moving objects or humans, referred to as \acrfull{so}, introduce uncontrolled, time varying effects, yielding a nonlinear system shaped jointly by RIS configurations and environmental evolution \cite{10077120}. It has been proven in \cite{11359994} that with adapative sensing accurate \acrshort{ue} localization can be achieved using large antenna arrays at the \acrshort{ue} and \acrshort{bs}, with the \acrshort{ris} acting as a virtual anchor in \acrshort{rses}.

In this work we consider the uplink localization problem in \acrshort{rses} in which a \acrshort{ue} repeatedly transmits the known pilot signals and \acrshort{bs} determines the location and orientation of the \acrshort{ue} via received pilots. Since the wireless channel depends on the \acrshort{ue} location, considered scenario can be viewed as an inverse mapping problem, where the goal is to infer the \acrshort{ue} location from the channel realizations, which are obtained by estimating the received pilot signals. The received signal at the \acrshort{bs} is reflected off multiple \acrshort{ris} and the scattering environment. The main idea is to exploit the multiple paths such that the \acrshort{bs} uses the received measurements and senses the environment via the dedicated sensing RIS elements and adaptively reconfigures the \acrshort{ris}, \acrshort{bs} and \acrshort{ue} beamforming vectors over multiple time slots, with an objective of localizing the \acrshort{ue} with maximum accuracy. The adaptive design approach is investigated in diverse multipath \acrshort{rse} scenarios and  the robustness and applicability of the approach across varying \acrshort{rses} conditions is evaluated. 

\subsection{Related Work}
Prior studies on \acrshort{ue} localization assisted by \acrshort{ris} can be broadly classified into model based and learning based approaches \cite{10858311}. Model based methods optimize the \acrshort{ris} configuration through analytical formulations, typically by minimizing the \acrfull{crlb} of the \acrshort{ue} position \cite{9500437, 9508872, 9148744, 9782100, 9774917}, providing useful theoretical performance bounds. These approaches rely on explicit relationships between the \acrshort{ue} location and measurements such as \acrfull{toa} or \acrshort{aoa} to construct the \acrfull{fim}. However, such formulations become impractical in multipath dominated \acrshort{rse}, where reliable path separation and association are difficult, particularly in narrowband indoor scenarios with unknown \acrshort{so}. Extensions including Bayesian \acrshort{fim} and robust \acrfull{bcrlb} incorporate prior uncertainty in channel or geometric parameters \cite{6541985}, and have enabled adaptive \acrshort{ris} beamforming schemes exploiting location uncertainty \cite{9772371}. Nevertheless, these methods often neglect scattering effects, limiting performance in dynamic environments. Recent works instead exploit multipath structure for improved estimation, including Bayesian multi target trajectory tracking \cite{11142591}, cooperative \acrshort{ris} assisted 3D positioning with low complexity estimator \cite{10884887}, and tensor based integrated sensing and communication framework supporting joint multi user communication and near and far field localization \cite{11200490}.

In contrast, model free localization techniques have gained increasing attention with advances in machine learning. Fingerprinting based method maps received signal measurements under multiple \acrshort{ris} configurations to \acrshort{ue} locations \cite{9548046}, while deep reinforcement learning has been applied to optimize \acrshort{ris} configurations for improved localization accuracy \cite{10054103}. Despite their effectiveness, these approaches typically rely on fixed or predefined configurations and therefore do not fully exploit adaptive \acrshort{ris} reconfiguration. Adaptive sensing has recently emerged as a powerful paradigm across wireless applications, including initial beam alignment in millimeter wave systems \cite{9724252,9746028,10124207}, precoder and combiner design for reciprocal \acrshort{mimo} channels \cite{10304519}, \acrshort{ris} assisted beam tracking \cite{10758383}, and localization in single path and multipath environments as well as with adaptive \acrshort{ris} codebooks \cite{10373816,10578020,10945967}. These works commonly employ \acrfull{rnn} based architectures that iteratively adapt beamforming using previously observed pilot signals, progressively refining sensing toward a desired objective. However, the resulting models are generally not directly generalizable to complex \acrshort{rses} \cite{11359994}. Recent studies address this limitation through learning driven approaches designed for \acrshort{rses}, including joint fingerprinting and feature extraction networks for 3D localization \cite{11411807}, self adaptive \acrshort{ris} optimization via learned surrogate models \cite{9860667}, RNN based codebook driven joint \acrshort{ris} phase and position estimation \cite{10765779}, and adaptive sensing strategies that iteratively refine \acrshort{ris} configurations using pilot feedback \cite{11359994}. Building on \cite{11359994}, this paper investigates adaptive \acrshort{bs} and \acrshort{ue} beamforming design in \acrshort{mimo} \acrshort{ris}-assisted \acrshort{rses} with randomly moving \acrshort{so}, targeting joint \acrshort{ue} location and orientation estimation. The proposed approach is evaluated under previously unexplored scenarios, including distributed \acrshort{ris} deployments, varying numbers of mobile scatterers, and variable sensing elements, and is analytically benchmarked against conventional \acrshort{bcrlb} based optimization methods.

\subsection{Main Contributions}
In this paper we develop an adaptive learning based localization approach for distributed \acrshort{ris} assisted \acrshort{mimo} networks operating in \acrshort{rses}, where both the propagation medium and network geometry are dynamic due to randomly moving \acrshort{so}. We propose a learning based approach to jointly design \acrshort{ris} configurations as well as \acrshort{bs} and \acrshort{ue} beamforming vectors with the objective of simultaneously estimating \acrshort{ue} position and orientation. Unlike conventional free-space or sparsely scattered localization methods that rely on \acrshort{aoa} extraction or geometric ray tracing, dynamic \acrshort{rses} exhibit dense and highly coupled multipath propagation, where localization information is embedded in complex spatio temporal scattering patterns. To address this challenge, we exploit sequential learning to capture long range dependencies in the received pilot measurements and environmental sensing measurements, enabling the network to learn the evolving relationship between scattering dynamics, \acrshort{ris} configurations, and localization performance. The proposed approach employs a \acrfull{bilstm}  based architecture to model temporal correlations induced by dynamic scattering, while adaptively refining \acrshort{ris} configuration, \acrshort{bs} and \acrshort{ue} beamforming vectors to progressively enhance localization accuracy. The main contributions of this work are summarized as follows.
\begin{enumerate}
\item In a distributed \acrshort{ris}-assisted uplink \acrshort{mimo} network operating in dynamic \acrshort{rse}, we study an adaptive sensing based localization approach for jointly and adaptively designing \acrshort{ris} configuration vectors, \acrshort{bs} beamforming vectors, and \acrshort{ue} beamforming vectors based on past measurements at \acrshort{bs} and sensed environmental scattering. The proposed approach enables simultaneous estimation of \acrshort{ue} position and orientation while accounting for scattering from randomly moving \acrshort{so}. 

\item We provide analytical benchmarking of the proposed learning based adaptive sensing approach using \acrshort{bcrlb} based modeling and optimization approach. This analysis quantifies the performance gap between the proposed data driven approach and conventional estimation theoretic limits in \acrshort{rses}.

\item We conduct a comprehensive performance evaluation of the proposed approach under a variety of \acrshort{rses} conditions, including different deployment configurations of distributed RIS arrays, varying numbers of randomly moving \acrshort{so}, and different numbers of sensing \acrshort{ris} elements. This study provides insights into the impact of environmental complexity and sensing capability on scattering estimation, localization performance and the robustness of proposed approach.

\item The proposed framework yields interpretable results, demonstrating that the joint sequential design of RIS configurations, and \acrshort{bs} and \acrshort{ue} beamforming vectors progressively enhances the quality of sensed multipath information and improves \acrshort{ue} localization and orientation estimation accuracy in \acrshort{rse}.
\end{enumerate}
Simulation results confirm that the proposed approach performs effectively in various \acrshort{rses} and  consistently outperforms adaptive schemes that do not explicitly account for scattering dynamics as well as the theoretical benchmark. 

The rest of the paper is organized as follows. The system model, end-to-end channel modeling, pilot transmission and observation protocol and the problem formulation are presented in Section \ref{systemmodel}. Section \ref{method} presents the devised neural network architecture. The analytical modeling approach is derived in Section \ref{analytical}. The simulation results are presented in Section \ref{results} and conclusions are drawn in Section \ref{conclusion}.

\textit{Notations}: Scalars, vectors, and matrices are represented in italic (e.g. \( b \)), lowercase bold letters (e.g. \( \mathbf{b} \)), and capital boldface letters (e.g. \( \mathbf{B} \)), respectively. Furthermore, \ \( |\cdot| \) denotes the modulus,  \( \hat{b} \) denotes an estimate of \( b \), $\text{diag}(\mathbf{y})$ presents a diagonal matrix with entries of $\mathbf{y}$ on diagonal.  The real and imaginary components of a complex value are given by \( \Re(\cdot) \) and \( \Im(\cdot) \), respectively. The expectation of a random variable is denoted by \( \mathbb{E}(\cdot) \). The complex Gaussian distribution with mean $m$ and variance $v$ is presented by $\mathcal{CN}(m,v)$.

\section{System Model and Problem Formulation}
\label{systemmodel}
\subsection{System Model}
\begin{figure}
    \centering
    \includegraphics[width=1\linewidth]{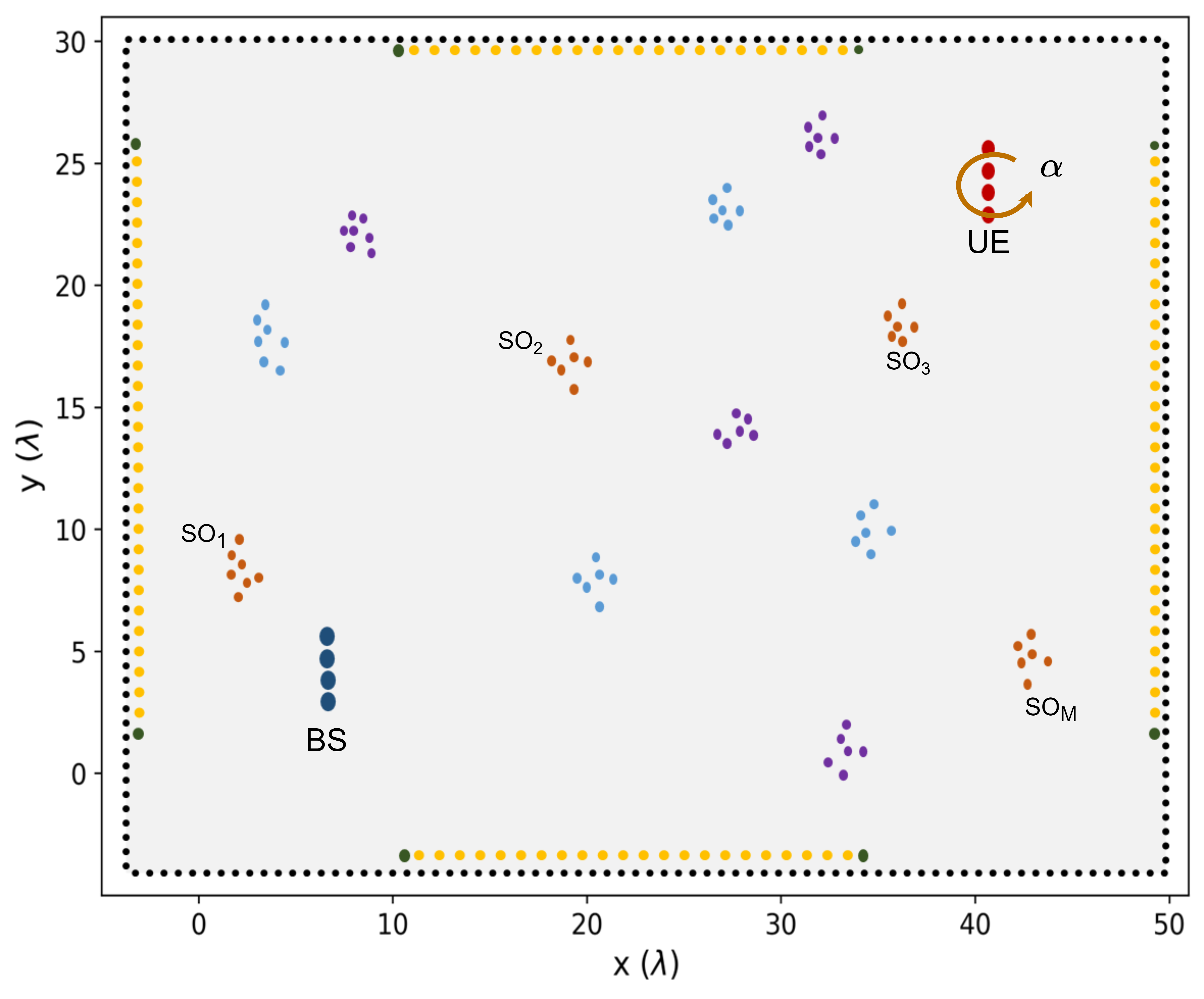}
    \caption{Depiction of the investigated \acrshort{rse}. The \acrshort{rse} contains $N_E$ dipoles. Their positions include a fixed black dipole fence forming the enclosure boundary and multiple clusters of colored dipoles located at different positions inside the enclosure, representing \acrshort{so}. The black dipoles remain stationary throughout the simulations, whereas the colored dipoles change position randomly in each realization. The positions of the dynamic dipoles that model \acrshort{so} are shown using three distinct colors for the first three realizations, namely plum, orange, and cyan, each corresponding to a different realization. A \acrshort{bs} with $N_{\text{BS}}$ antennas is illustrated by blue dipoles, \acrshort{ue} with $N_{\text{U}}$ antennas is shown as red dipoles, and a distributed uniform linear \acrshort{ris} configuration consisting of $N_{\text{RIS}}$ elements represented by yellow dipoles that partially cover the four enclosure walls formed by the black dipole fence. The subset $S_{\text{RIS}}$ of \acrshort{ris} elements used for environment sensing is indicated by green dipoles.} 
    \label{fig1}
\end{figure}

Consider the problem of uplink \acrshort{ue} localization in an indoor \acrshort{rse} comprising distributed \acrshort{ris} installation, a multi-antenna \acrshort{ue}, a multi-antenna \acrshort{bs} and $M$ \acrshort{so}. The \acrshort{bs}, at known location, aims to estimate the location $\mathbf{u} = [u_x,\, u_y]^{\mathrm{T}} \in \mathbb{R}^2$ and orientation \(\alpha \in [0, 2\pi) \) of \acrshort{ue} with the assistance of a uniform linear array of \acrshort{ris} distributed installation on the periphery of the communication environment,  as shown in Figure \ref{fig1} (top view). \acrshort{rse} is modeled using the physics based end-to-end channel model \cite{9856592}, where the environment is treated as a regularly shaped enclosure whose dimensions are large relative to the wavelength scale $\lambda$.

The \acrshort{ue} broadcasts pilot signals and \acrshort{bs} can adapt the received pilots by configuring the phase shifts of the \acrshort{ris} elements via the \acrshort{ris} controller through the control signals. We assume that $M$ \acrshort{so} are present in \acrshort{rse} and their movement in the periphery is random, mimicking freely moving objects in any indoor environment. The signals from \acrshort{ue} experience multiple reflections and scatter off walls, multiple \acrshort{ris} elements, and \acrshort{so} in the propagation environment. This scattering and the movement of \acrshort{so} further alter the propagation path and create a dynamic and complex fading wireless communication environment \cite{9475155}.

We consider uplink transmission from a multi-antenna \acrshort{ue} with $N_{\rm UE}$ antennas and a single RF chain to a BS with $N_{\rm BS}$ antennas. Let 
$\mathbf{\omega_{\text{BS}}} \in \mathbb{C}^{N_{\text{BS}} \times 1}$  and $\mathbf{\omega_{\text{UE}}} \in \mathbb{C}^{N_{\text{UE}} \times 1}$
denote the beamforming vectors at \acrshort{bs} and \acrshort{ue}, respectively,  
whose elements satisfy a unit-modulus constraint, i.e., 
\([\mathbf{\omega}]_{q} = e^{j \omega_{q}}\) with \(\omega_{q} \in [0, 2\pi)\). Let \(N_{\text{RIS}}\) be the number of reflective elements at \acrshort{ris}, each applying a tunable phase shift \(\delta_n \in [0, 2\pi)\) to the incident signal. Let \acrshort{ris} configuration  vector be \(\mathbf{k} \in \mathbb{C}^{1 \times N_{\text{RIS}}}\), where each entry satisfies the unit-modulus constraint, i.e., \(|[\mathbf{k}]_n| = 1, \forall n \in \{1, \dots, N_{\text{RIS}}\}\). A subset of RIS elements \(S_{\text{RIS}}\) with the power splitting ratio set to zero to disable signal reflection is placed at fixed locations along the periphery of the environment for channel sensing in order to localize \acrshort{so} \cite{10352433, 10077120}. The estimated location of $M$ \acrshort{so} are represented by the vector $\mathbf{p}={[\mathbf{p_1}, \mathbf{p_2},... ,\mathbf{p_M}]} \in \mathbb{R}^{2M}$. This allows the scattering environment sensing without requiring any cooperation from the \acrshort{ue}.  

\subsection{End-to-End Channel Modeling}
We assume the channel is constant over a coherence interval spanning multiple time frames, with small-scale \acrshort{rse} perturbations, such as slight \acrshort{so} movements or multipath changes, modeled as discrete-time structured fast-fading to capture temporal correlations and enhance scattering and \acrshort{ue} localization. We consider an indoor \acrshort{rse} with low \acrshort{ue} motion with coherence interval exceeding the sensing interval. Unlike the linear \acrshort{bs}–\acrshort{ris}–\acrshort{ue} cascade in free space, \acrshort{rse} involves complex multipath with multiple RIS elements and dynamic \acrshort{so}, rendering the channel highly nonlinear and analytically intractable \cite{9475155}. We employ a physics-based end-to-end channel for \acrshort{ris}-parameterized \acrshort{rse} \cite{9856592}, modeling wireless entities as configurable dipole assemblies. This approach realistically captures scattering effects while adhering to fundamental electromagnetic principles, including space-time causality, frequency dispersion, RIS phase-amplitude coupling, mutual coupling, and long-range correlations.

As shown in Figure \ref{fig1}, the end-to-end channel models \acrshort{so}, walls, \acrshort{ris}, and transceivers as dipole arrays. The total number of dipoles is $N = N_\text{BS} + N_\text{UE} + N_\text{E} + N_\text{RIS}$. Let \( \textbf{H}_{\text{BS}-\text{U}} \in \mathbb{C}^{N_{\text{UE}} \times N_{\text{BS}}} \) denote the stochastic channel frequency response at $U$ be given by
\begin{equation}
\mathbf{H}_{\text{BS}-\text{U}} = \mathbf{V}\left[(N_{\text{BS}} + 1) : (N_{\text{BS}} + N_{\text{U}}), 1 : N_{\text{BS}} \right],   
\end{equation}
where
\begin{equation}
\mathbf{V} \triangleq \text{diag} \left( \left[ \alpha^{-1}_1, \alpha^{-1}_2, \dots, \alpha^{-1}_N \right] \right) \mathbf{W}^{-1},
\end{equation}
The \( i^{th} \) diagonal element of \( N \times N \) complex-valued matrix \(\mathbf W \) represents the inverse polarizability \( \alpha^{-1}_i (f) \) of the \( i^{th} \) dipole.  The polarizability \( \alpha_i (f) \) of the \( i \)th dipole is given by
\begin{equation}
\alpha_i(f) = \frac{\chi^2_i}{4\pi^2 f^2_{\text{res},i} - 4\pi^2 f^2 + j(\gamma^R_i + 2\pi f \Gamma^L_i)},
\end{equation}
where \( \chi^2_i \) scales the amplitude of \( \alpha_i(f) \), \( f_{\text{res},i} \) is the resonance frequency, \( \gamma^R_i \) captures radiation damping and, \( \Gamma^L_i \geq 0 \) represents absorptive damping.

The off-diagonal entry \( (i, j) \) of \(\mathbf W \) is \( -G_{ij} (f) \), representing the negative of two-dimensional free-space Green’s function between dipoles at positions $\mathbf{r}_i$ and $\mathbf{r}_j$ is given by
\begin{equation}
G_{ij} (\mathbf{r}_i, \mathbf{r}_j, f) \triangleq -j \frac{k^2}{4\epsilon\delta} \text{H}^{(2)}_0 \left( k |r_i - r_j| \right).
\end{equation}
where \( \text{H}^{(2)}_0 (\cdot) \) is Hankel function of the second kind.

\subsection{Pilot Transmission and Reception}
Assuming a narrowband system model, when a localization request is initiated the \acrshort{ue} transmits a sequence of \( T \) known uplink  pilot symbols, with beamforming $\mathbf{\omega_{\text{UE}}}^{(t)}$,  to the \acrshort{bs} over \( T \) time frames. Let \( \textbf{x}^{(t)} \in \mathbb{C}^{N_{U} \times 1} \) denote the pilot sent at the $t$-th frame. At the \acrshort{bs}, the received signal from the \acrshort{ue} is combined using a beamforming vector $\mathbf{\omega_{\text{BS}}}^{(t)}$. The effective signal received by the \acrshort{bs}, 
\(\mathbf{y}^{(t)} \in \mathbb{C}^{N_{U} \times 1}\), 
consisting of contributions from the direct transmission, if present, 
as well as multipath components resulting from reflections off the \acrshort{ris} and the surrounding scattering environment, is given by 
\begin{equation}
\mathbf{y}^{(t)}= (\omega_{\rm{BS}}^{(t)})^\top\textbf{H}_{\text{BS}-\text{U}}^{(t)}(\omega_{\rm{UE}}^{(t)})^\top\mathbf{k}^{(t)}  \textbf{x}^{(t)} + \textbf{w}^{(t)}, \quad t \in \mathbb{Z}
\end{equation}
where \( \textbf{w}^{(t)} \sim  \mathcal{CN}(0,\sigma^2 )\) is additive Gaussian noise. The received pilot signal is the function of \acrshort{ue} and \acrshort{bs} beamforming vectors as well as the \acrshort{ris} configuration at the $t$-th time stage. 

In addition, the \acrshort{bs} uses \( S_{\text{RIS}} \) \acrshort{ris} elements to sense the environment and estimate the location of scattering objects \( \mathbf{p}^{\rm{(t)}} \) at \( t \)-th frame by measuring the channel $\mathbf{H}_{\text{BS}-\text{S}} \in \mathbb{C}^{S_{\text{RIS}} \times N_{\text{BS}}}$. The observations are given by
\begin{equation}
\mathbf{y}_s^{(t)} = \mathbf{H}_{\text{BS}-\text{S}}^{(t)} \mathbf{x}^{(t)} + \mathbf{w}^{(t)}, \quad t \in \mathbb{Z}.
\end{equation}  

The goal of this paper is to analyse various \acrshort{rses} with adaptive joint optimization of the \acrshort{bs} beamforming vector, \acrshort{ue} beamforming vector and and \acrshort{ris} configurations to improve the \acrshort{ue} location and orientation estimate in addition to estimation of scattering in the environment.

\subsection{Problem Formulation}
The main objective in this work is to estimate the unknown \acrshort{ue} location \( \mathbf{u} \) and orientation $\alpha$ from $T$ received observations $[\mathbf{y}^{(t)}]_{t=0}^{T-1}$, given known \acrshort{bs} and \acrshort{ris} locations and a predicted \acrshort{so} location at $t$-th time frame. The \acrshort{so} state $\mathbf{p}$ is inferred from \acrshort{ris} sensing signals $[\mathbf{y}_s^{(\tau)}]_{\tau=0}^{t}$ at $S_{\text{RIS}}$ elements to model scattering in the \acrshort{rse}. In this setting, \acrshort{ris} provides directional information by acting as virtual anchors through its dense reflective elements. We, therefore, propose to design the \acrshort{ris} configuration, \acrshort{bs} beamforming vector and \acrshort{ue} beamforming vector over $t$ time stages to improve the \acrshort{ue} location and orientation estimates at $T$. 

The field observed at any spatial point in \acrshort{rse} is formed by the superposition of multiple propagation signals, with fluctuations caused by dynamic \acrshort{so} whose behavior depends on their position $\mathbf{p}^{(t)}$. Define the mapping \( G^{(t)}(\cdot):\mathbb{C}^{t+1} \to \mathbb{R}^{2M}\) from the received pilot signals at \( S_{\text{RIS}} \) to estimated location of \acrshort{so} for time frame $t$ as
\begin{equation}
\label{pt}
    \hat{\mathbf{p}}^{(t)} = G^{(t)}([\textbf{y}_s^{(\tau)}]_{\tau=0}^{t}]),
\end{equation}
where $\tau$ measures previous time steps up to the current time $t$.

\begin{figure*}[t!]
    \centering   \includegraphics[width=\linewidth]{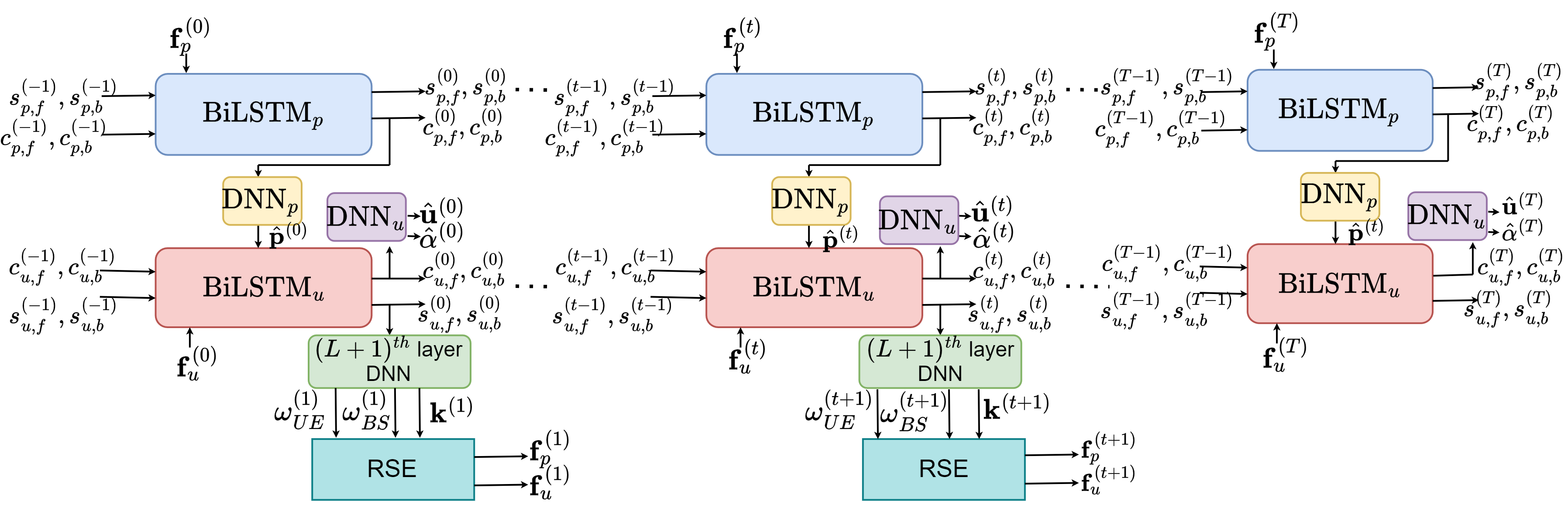}
    \caption{Block diagram of the proposed \acrshort{ue} localization approach with adaptive sensing in RIS-assisted \acrshort{rse}. }
    \label{nnarch}
\end{figure*}

Define the mapping $F^{(t)}: \mathbb{C}^{t+1} \times \mathbb{R}^{2M}\to  \mathbb{C}^{N_{\text{BS}} \times 1}  \times \mathbb{C}^{N_{\text{UE}} \times 1}  \times \mathbb{C}^{1 \times N_{\text{RIS}}}  $ from the previously received pilot signals at \acrshort{bs} and mapping in \eqref{pt} to jointly design the next \acrshort{ris} configuration, \acrshort{bs} beamforming vector and \acrshort{ue} beamforming vector for the $(t+1)$-th stage as 
\begin{equation}
\label{F}
[\omega_{\rm{BS}}^{(t+1)}, \omega_{\rm{UE}}^{(t+1)}, \textbf{k}^{(t+1)}] = F^{(t)}\left( [\textbf{y}^{(\tau)}]_{\tau=0}^{t}, \hat{\mathbf{p}}^{(t)} \right).
\end{equation}

Define the mapping $Q: \mathbb{C}^T \times \mathbb{R}^{2M} \to \mathbb{R}^3$ from all pilot signals received in $T$ time frames and estimated scattering to the final \acrshort{ue} location $\hat{\mathbf{u}}$ and orientation estimate $\hat{\alpha}$ as
\begin{equation}
\label{Q}
[\hat{\mathbf{u}}, \hat{\alpha}] = Q\left( [\textbf{y}^{(t)}]_{t=0}^{T-1},  \hat{\mathbf{p}}^{(t)}\right). 
\end{equation}

The ultimate objective of \acrshort{ris} aided localization is to minimize the localization error based on the received pilot signals. This objective can be expressed as an optimization problem that seeks to minimize the localization error, and is formulated as
\begin{subequations} \label{eq:optimization_problem}
\begin{align}
\min_{\{G^{(t)}(\cdot), F^{(t)}(\cdot)\}_{t=0}^{T-1}, Q(\cdot)} & \bigg[\sqrt{\mathbb{E} \left[ \|\hat{\mathbf{u}} - \mathbf{u}\|^2 \right]}, \sqrt{\mathbb{E} \left[ \|\hat\alpha - \alpha\|^2 \right]}\bigg] \label{eq:optimization_objective} \\
\text{subject to}
 \label{9b}\quad \hat{\mathbf{p}}^{(t)} &= G^{(t)}([\textbf{y}_s^{(\tau)}]_{\tau=0}^{t}),\\
 [\omega_{\rm{BS}}^{(t+1)}, \omega_{\rm{UE}}^{(t+1)}, \textbf{k}^{(t+1)}]  &= F^{(t)}\left( [\textbf{y}^{(\tau)}]_{\tau=0}^{t}, \hat{\mathbf{p}}^{(t)} \right), \label{9c} \\
 [\hat{\mathbf{u}}, \hat{\alpha}] &= Q\left( [\textbf{y}(t)]_{t=0}^{T-1}, \hat{\mathbf{p}}^{(t)} \right), \label{9d}\\
 \left| \left[ \mathbf{k}^{(t+1)} \right]_n \right| &= 1, \quad \forall n \in N_{RIS}, t, \label{9e}\\
 \left| \left[ \mathbf{\omega_{\rm{BS}}}^{(t+1)} \right]_n \right| &= 1, \quad \forall n \in N_{\rm{BS}}, t, \label{9f}\\
 \left| \left[ \mathbf{\omega_{\rm{UE}}}^{(t+1)} \right]_n \right| &= 1, \quad \forall n \in N_{\rm{BS}}, t. \label{9g}
\end{align}
\end{subequations}

The analytical solution to the problem in \eqref{eq:optimization_problem} is difficult since it involves complex mappings for the estimation of \acrshort{ue} location and orientation alongside scattering estimation, with joint optimized design of \acrshort{ris} configuration, \acrshort{bs} and \acrshort{ue} beamforming vectors. A common approach is to address the problem in a simpler yet greedy manner by reducing the estimation \acrshort{rmse} at every time step. However, evaluating the exact objective in \eqref{eq:optimization_problem} requires high dimensional integration, which is difficult to compute numerically. As a result, earlier studies \cite{6541985} suggest \acrshort{bcrlb} as a surrogate metric to be minimized at each time step. This approach has been used as a benchmark for comparison with the proposed \acrshort{bilstm} based data driven approach. 
We employ a \acrshort{bilstm} network to perform the mapping in \eqref{pt} to perform scattering estimation and for mapping $F^{(t)}(.)$, to construct the \acrshort{ris} configuration, \acrshort{bs} and \acrshort{ue} beamforming vectors. The same network is used to adaptively select the set of next \acrshort{ris} configuration vector, \acrshort{bs} beamforming vector as well as the \acrshort{ue} beamforming vector for subsequent measurement, thus, solving the problem stated in \eqref{eq:optimization_problem}. Our learning based approach can achieve substantially better performance than \acrshort{bcrlb} based design, as demonstrated by the simulation results in Section \ref{results}.

\section{Proposed Joint User Location and Orientation Estimation with Environment Sensing Solution}
\label{method}
The \acrshort{ue} localization in \acrshort{rse} with \acrshort{ris} has two main stages: sensing the environment to gain context and determining optimal \acrshort{bs} and \acrshort{ue} beamforming vector, and \acrshort{ris} configurations for \acrshort{ue} localization. Sensing estimates the \acrshort{so} state from field measurements, guiding beamforming and configuration design for accurate localization. 
We propose a learning based adaptive approach that jointly designs the \acrshort{ris} configuration and the \acrshort{bs} and \acrshort{ue} beamforming to solve the localization problem in \eqref{eq:optimization_problem}. It involves offline training specific to the \acrshort{rse} and online inference during deployment. At each time step $t$, the \acrshort{ris} configuration $ \mathbf{k}^{(t+1)} $, the \acrshort{bs} beamforming $ \mathbf{\omega}_{\mathrm{BS}}^{(t+1)} $, and the \acrshort{ue} beamforming $ \mathbf{\omega}_{\mathrm{UE}}^{(t+1)} $ vectors are chosen based on the received signal measurements history up to $t$, $ [\mathbf{y}^{(\tau)}]_{\tau=0}^{t} $ and the current \acrshort{so} estimate $ \hat{\mathbf{p}}^{(t)} $. Since the number of measurements increases over time, a neural network is used to compress past observations into a fixed dimensional representation that preserves the information needed for efficient adaptation.

We use a \acrshort{bilstm} network to model temporal correlations and encode sequential measurements efficiently. Instead of adapting the \acrshort{ris} configurations and beamforming vectors only from  \acrshort{bs} pilots measurements, the proposed two stage framework consists of scattering estimation \acrshort{bilstm}, which estimates the time varying \acrshort{so} positions from sensing measurements at $ S_{\text{RIS}} $ to characterize the \acrshort{rse}, and
\acrshort{ue} localization \acrshort{bilstm}, which designs the \acrshort{ris} configuration, \acrshort{bs} beamforming vector, and the \acrshort{ue} beamforming vector for time frame $(t+1)$ using the estimated scattering and the \acrshort{ue} pilot at time frame $( t )$ to enable accurate localization at the final time frame $( T )$. The neural network architecture is demonstrated in Figure \ref{nnarch}. 

\subsection{Scattering Estimation BiLSTM}
At each time step \( t \), a \acrshort{bilstm} processes the sensing feature set that concatenates the magnitude and phase of sensed signal $\mathbf{f}_p^{(t)} = [\, |\mathbf{y}_s^{(t)}| , \angle \mathbf{y}_s^{(t)} \,]$
and updates the forward and backward hidden state vectors \( \mathbf{s}_{p,f}^{(t)} \) and \( \mathbf{s}_{p,b}^{(t)} \) and forward and backward cell state vectors \( \mathbf{c}_{p,f}^{(t)} \) and \( \mathbf{c}_{p,b}^{(t)} \) as 
\begin{equation}
\label{statesupdate}
(\mathbf{c}_{p,f}^{(t)}, \mathbf{c}_{p,b}^{(t)}, \mathbf{s}_{p,f}^{(t)},  \mathbf{s}_{p,b}^{(t)}) = \mathrm{BiLSTM}_p\big(\mathbf{f}_p^{(t)}, \mathbf{c}_{p,f}^{(t-1)}, \mathbf{c}_{p,b}^{(t-1)}, \mathbf{s}_{p,f}^{(t-1)},  \mathbf{s}_{p,b}^{(t-1)}\big),
\end{equation}
where $\mathbf{c}_{p,f}^{(t-1)}, \mathbf{c}_{p,b}^{(t-1)} $ are forward and backward cell state vectors, $\mathbf{s}_{p,f}^{(t-1)},  \mathbf{s}_{p,b}^{(t-1)}$ are forward and backward hidden state vectors that correspond to the information based on temporal measurements before the $t$-th time step. The update rule for $\mathrm{BiLSTM}(.)_1$ is as discussed in \cite{11359994}.

The forward and backward states jointly capture temporal dependencies from both past and future sensing samples, providing an accurate representation of the evolving scattering environment. The concatenated cell states are then fed to a fully connected output layer to estimate the \acrshort{so} position as
\begin{equation}
\label{soestimate}
\hat{\mathbf{p}}^{(t)} = \text{DNN}_{\mathrm{p}}\!\left([\mathbf{c}_{p,f}^{(t)} ; \mathbf{c}_{p,b}^{(t)}]\right),
\end{equation}
where \( \text{DNN}_{\mathrm{p}}(\cdot) \) is a trainable neural network. This estimate directly corresponds to the mapping in \eqref{9b}.

\begin{algorithm}[t]
\LinesNumbered
\caption{Joint \acrshort{ue} Location and Orientation Estimation with Environmental Sensing in \acrshort{rse} Algorithm}
\label{algo:bilstm_ris}
\KwIn{Initialize RIS configuration $\textbf{k}^{(0)}$ , \newline 
Initialize \acrshort{bs} beamforming vector ${\boldsymbol{\omega}}_{\mathrm{BS}}^{(0)}$, \newline 
Initialize \acrshort{ue} beamforming vectors ${\boldsymbol{\omega}}_{\mathrm{BS}}^{(0)}$.}

\KwOut{Optimized RIS configuration ${\textbf{k}}_{\mathrm{BS}}^{(t+1)}$,\newline 
Optimized BS beamforming vector ${\boldsymbol{\omega}}_{\mathrm{BS}}^{(t+1)}$, \newline 
Optimized \acrshort{ue} beamforming vectors ${\boldsymbol{\omega}}_{\mathrm{BS}}^{(t+1)}$, \newline 
\acrshort{ue} location estimate $\hat{\mathbf{u}}^{(T)}$, \newline 
\acrshort{ue} orientation estimate $\hat{\alpha}^{(T)}$, \newline 
Optimized $\text{\acrshort{bilstm}}_p(.)$, $\text{\acrshort{bilstm}}_u(.)$,  $\text{DNN}_{\mathrm{p}}, \text{DNN}_{\mathrm{u}}$ network.  }

\For{$t=1$ \KwTo $T$}{
    \acrshort{ue} transmits pilot signal $x^{(t)}$\;
    
    \acrshort{bs} receives $y^{(t)}$ and $S_{\text{RIS}}$ receive $y_s^{(t)}$\;
    
    Update state vectors $\mathbf{s}_{p,f}^{(t)},\mathbf{s}_{p,b}^{(t)},\mathbf{c}_{p,f}^{(t)},\mathbf{c}_{p,b}^{(t)}$ using \eqref{statesupdate}\;
    
    Estimate $\hat{\mathbf{p}}^{(t)}$ using \eqref{soestimate}\;

    Estimate \acrshort{rmse} loss $L_p$ in \eqref{lossp}\;
    
    Update state vectors $\mathbf{s}_{u,f}^{(t)},\mathbf{s}_{u,b}^{(t)},\mathbf{c}_{u,f}^{(t)},\mathbf{c}_{u,b}^{(t)}$ using \eqref{update}\;
    
    Design $\bar{\mathbf{k}}^{(t+1)},\bar{\boldsymbol{\omega}}_{\mathrm{BS}}^{(t+1)},\bar{\boldsymbol{\omega}}_{\mathrm{UE}}^{(t+1)}$ using \eqref{vectorsdesign}\;

    Normalize $\bar{\mathbf{k}}^{(t+1)},\bar{\boldsymbol{\omega}}_{\mathrm{BS}}^{(t+1)},\bar{\boldsymbol{\omega}}_{\mathrm{UE}}^{(t+1)}$ using \eqref{risconfig}\;
}

Estimate \acrshort{ue} position $\hat{\mathbf{u}}^{(T)}$ and orientation $\hat{\alpha}^{(T)}$ using \eqref{estimuser_alpha}\;
Compute \acrshort{rmse} loss $L_u$ and $L_\alpha$ using \eqref{lossu}\;

\Return ${\mathbf{k}}^{(t+1)}, {\boldsymbol{\omega}}_{\mathrm{BS}}^{(t+1)}, {\boldsymbol{\omega}}_{\mathrm{UE}}^{(t+1)}, \hat{\mathbf{u}}^{(T)}, \hat{\alpha}^{(T)}$, $\text{\acrshort{bilstm}}_p(.)$, $\text{\acrshort{bilstm}}_u(.)$,  $\text{DNN}_{\mathrm{p}}, \text{DNN}_{\mathrm{u}}$.
\end{algorithm}

\subsection{UE Localization BiLSTM}
At each time step \( t \), a \acrshort{bilstm} updates its forward and backward hidden states, \( \mathbf{s}_{u,f}^{(t)}\) and \( \mathbf{s}_{u,b}^{(t)}\), along with the corresponding cell states \( \mathbf{c}_{u,f}^{(t)}\) and \(\mathbf{c}_{u,b}^{(t)}\), based on the input \(\mathbf{f}_u^{(t)} = [\, |\mathbf{y}^{(t)}|, \angle \mathbf{y}^{(t)}, \hat{\mathbf{p}}^{(t)} \,]\).
\begin{equation}
\label{update}
(\mathbf{c}_{u,f}^{(t)}, \mathbf{c}_{u,b}^{(t)}, \mathbf{s}_{u,f}^{(t)},  \mathbf{s}_{u,b}^{(t)}) = \\\mathrm{BiLSTM}_u\big(\mathbf{f}_u^{(t)}, \mathbf{c}_{u,f}^{(t-1)}, \mathbf{c}_{u,b}^{(t-1)}, \mathbf{s}_{u,f}^{(t-1)},  \mathbf{s}_{u,b}^{(t-1)}\big),
\end{equation}
The hidden states and cell state vector updates follow the same rules as in \cite{11359994}.

The combined hidden state vector $s_u^{(t)} = [\, s_{u,f}^{(t)} ; s_{u,b}^{(t)}] \,$ is used to design the \acrshort{ris} configuration, \acrshort{bs} beamforming and \acrshort{ue} beamforming vectors, respectively. It is passed through a fully connected \acrfull{dnn} with \( L_n \) layers to produce an intermediate representation
\begin{equation}
\label{gamma_alpha}
\gamma^{(t)} = \beta_{L_n} \Big( \mathbf{W}_{L_n} \beta_{L_n-1} ( \dots \beta_1 ( \mathbf{W}_1 s_u^{(t)} + \mathbf{b}_1 ) \dots ) + \mathbf{b}_{L_n} \Big),
\end{equation}
where \( \beta_l(x) = \max(0, x) \) is the ReLU activation and \( \mathbf{W}_l, \mathbf{b}_l \) are trainable parameters.

The three \((L+1)\)-th layers map \(\gamma(t)\) into the RIS configuration vector, BS beamforming vector and UE beamforming vector as follows
\begin{subequations}
\label{vectorsdesign}
\begin{align}
\bar{\mathbf{k}}^{(t+1)} &= \dot{\mathbf{W}}_{L_n+1} \gamma^{(t)} + \dot{\mathbf{b}}_{L_n+1}, \label{RIS_vector}\\
\bar{\mathbf{\omega}}_\mathrm{BS}^{(t+1)} &= \ddot{\mathbf{W}}_{L_n+1} \gamma^{(t)} + \ddot{\mathbf{b}}_{L_n+1}, \label{BS_vector}\\
\bar{\mathbf{\omega}}_\mathrm{UE}^{(t+1)} &= \dddot{\mathbf{W}}_{L_n+1} \gamma^{(t)} + \dddot{\mathbf{b}}_{L_n+1}, \label{UE_vector}
\end{align}
\end{subequations}
where $\bar{\mathbf{k}}^{(t+1)}$ has the sequence of real and imaginary components of the \acrshort{ris} configuration, while \(\bar{\mathbf{\omega}}_\mathrm{BS}^{(t+1)}\) and \(\bar{\mathbf{\omega}}_\mathrm{UE}^{(t+1)}\) contain the real and imaginary parts of the \acrshort{bs} and \acrshort{ue} beamforming vectors. The dimensions of the trainable weights and biases are set to ensure correct output sizes: \(\bar{\mathbf{k}}^{(t+1)} \in \mathbb{R}^{2 N_\mathrm{RIS}}\), \(\bar{\mathbf{\omega}}_\mathrm{BS}^{(t+1)} \in \mathbb{R}^{2 N_\mathrm{BS}}\), and \(\bar{\mathbf{\omega}}_\mathrm{UE}^{(t+1)} \in \mathbb{R}^{2 N_\mathrm{UE}}\).
Each element is normalized to satisfy the unit-modulus constraint, 
\begin{subequations}
\label{risconfig}
\begin{align}
[\mathbf{k}^{(t+1)}]_n =
\frac{[\Re(\bar{\mathbf{k}}^{(t+1)})]_n}{\sqrt{[\Re(\bar{\mathbf{k}}^{(t+1)})]_n^2 + [\Im(\bar{\mathbf{k}}^{(t+1)})]_n^2}}\\ \nonumber
+ j \frac{[\Im(\bar{\mathbf{k}}^{(t+1)})]_n}{\sqrt{[\Re(\bar{\mathbf{k}}^{(t+1)})]_n^2 + \Im[(\bar{\mathbf{k}}^{(t+1)})]_n^2}},\\
[{\mathbf{\omega}}_\mathrm{BS}^{(t+1)}]_q =
\frac{[\Re(\bar{\mathbf{\omega}}_\mathrm{BS}^{(t+1)})]_q}{\sqrt{[\Re(\bar{\mathbf{\omega}}_\mathrm{BS}^{(t+1)})]_q^2 + [\Im(\bar{\mathbf{\omega}}_\mathrm{BS}^{(t+1)})]_q^2}}\\ \nonumber
+ j \frac{[\Im(\bar{\mathbf{\omega}}_\mathrm{BS}^{(t+1)})]_q}{\sqrt{[\Re(\bar{\mathbf{\omega}}_\mathrm{BS}^{(t+1)})]_q^2 + \Im[(\bar{\mathbf{\omega}}_\mathrm{BS}^{(t+1)})]_q^2}},\\
[{\mathbf{\omega}}_\mathrm{UE}^{(t+1)}]_q =
\frac{[\Re(\bar{\mathbf{\omega}}_\mathrm{UE}^{(t+1)})]_q}{\sqrt{[\Re(\bar{\mathbf{\omega}}_\mathrm{UE}^{(t+1)})]_q^2 + [\Im(\bar{\mathbf{\omega}}_\mathrm{UE}^{(t+1)})]_q^2}}\\ \nonumber
+ j \frac{[\Im(\bar{\mathbf{\omega}}_\mathrm{UE}^{(t+1)})]_q}{\sqrt{[\Re(\bar{\mathbf{\omega}}_\mathrm{UE}^{(t+1)})]_q^2 + \Im[(\bar{\mathbf{\omega}}_\mathrm{UE}^{(t+1)})]_q^2}}. 
\end{align}
\end{subequations}
This design satisfies the mapping in \eqref{9c}, \eqref{9e}, \eqref{9f} and \eqref{9g}.

Finally, the network estimates both the \acrshort{ue} location and orientation at time \( T \) via
\begin{equation}
\label{estimuser_alpha}
\begin{bmatrix}
\hat{\mathbf{u}}^{(T)} \\ \hat{\alpha}^{(T)}
\end{bmatrix}
= \text{DNN}_u \big( [\, \mathbf{c}_{u,f}^{(T)} ; \mathbf{c}_{u,b}^{(T)} \,] \big),
\end{equation}
where \( \text{DNN}_u(\cdot) \) is a fully connected network mapping the concatenated forward and backward cell states to the final UE position \(\hat{\mathbf{u}}^{(T)}\) and orientation \(\hat{\alpha}^{(T)}\) estimates, as stated in \eqref{eq:optimization_objective}.

\subsection{Loss Function}
The composite loss function for the overall network is based on the \acrshort{rmse} loss of \acrshort{ue} location estimation $L_u$ and orientation estimation $L_{\alpha}$ as well as the \acrshort{so} estimation loss $L_p$ given as

\begin{subequations}
\label{lossu}
\begin{align}
L_u=\sqrt{\mathbb{E}\!\left[\|\hat{\mathbf{u}}^{(T)}-\mathbf{u}\|^2\right]} \\
L_\alpha=\sqrt{\mathbb{E}\!\left[\|\hat{\boldsymbol{\alpha}}^{(T)}-\boldsymbol{\alpha}\|^2\right]},
\end{align}
\end{subequations}
\begin{equation}
\label{lossp}
L_p=\sqrt{\mathbb{E}\!\left[\|\hat{\mathbf{p}}^{(T)}-\mathbf{p}\|^2\right]}.
\end{equation}
These losses guides the network to accurately localize the \acrshort{ue} while adapting the \acrshort{ris} configuration, \acrshort{bs} and \acrshort{ue} beamforming vectors to changing propagation conditions. It promotes the design of sequence of \(T\) aforementioned vectors that minimize the final localization error. After training, the model generates them directly from the received signals for any \acrshort{ue} and \(S_{\text{RIS}}\) across the coverage region. Since only the estimate at time \(T\) is penalized, the network is free to explore different vectors designs during earlier stages to achieve the best final accuracy. The algorithm is summarized in Algorithm \ref{algo:bilstm_ris}.

\section{Analytical Modeling of UE localization with Environment Sensing in RSE}
\label{analytical}
Analytical modeling of the \acrshort{ue} localization problem with scattering estimation in \acrshort{rse} is a valuable benchmark. The proposed learning driven approach addresses \acrshort{ue} location and orientation estimation problem in \eqref{eq:optimization_problem} by estimating scattering in \acrshort{rse} in \eqref{pt} and jointly optimizing the \acrshort{ris} configuration, \acrshort{bs} and \acrshort{ue} beamforming vectors mapping $\{F^{(t)}(\cdot)\}$ in \eqref{F} over $t$ time frames, with the goal of reducing localization error. Computational complexity of an analytical optimization of this multistage design would grow as the number of time frames increases. For this reason, analytical modeling of this formulation would need stage by stage greedy design of $F^{(t)}(\cdot)$. In addition, evaluating the exact \acrshort{rmse} is analytically intractable, thus, such modeling approach would rely on optimizing a tractable lower bound on \acrshort{rmse} instead.

In this section, we outline an analytical modeling approach for \acrshort{ue} localization in \acrshort{rse}. The approach is devised on similar lines as in \cite{6541985, 10373816}. Rather than learning a direct mapping from observations to \acrshort{ris} configuration, the mapping function in $F^{(t)}(\cdot)$ is decomposed into two consecutive mappings: first from the accumulated observations to the \acrshort{bcrlb}, and then from that bound to the \acrshort{ris} configuration vector. At every time step, the posterior distribution of the UE location is refined using previously received pilots. This updated posterior is then used to recompute the conditional BCRLB. The \acrshort{ris} configuration vector is finally selected so as to minimize this conditional \acrshort{bcrlb}.

\subsection{Problem Formulation}
The unknown \acrshort{ue} location parameters to be estimated are presented here by grouping the position coordinates and orientation into a single vector given by
\begin{equation}
 \mathbf{u}_e = \begin{bmatrix} x_u \quad y_u \quad  \alpha\end{bmatrix}^\top \in \mathbb{R}^3.
\end{equation}
For simplicity the vector $\mathbf{u}_e$ provides a compact representation of the \acrshort{ue} spatial state and enables unified estimation from received pilot signal measurements. Let the \acrshort{so} locations be given as
\begin{equation}
\mathbf{p} = 
\begin{bmatrix}
\mathbf{p}_1^\top, \dots, \mathbf{p}_M^\top
\end{bmatrix}^\top 
\in \mathbb{R}^{2M}.
\end{equation}

We assume both $\mathbf{u}_e$ and $\mathbf{p}$ are random vectors with known priors $ f(\mathbf{u}_e)$ and $f(\mathbf{p})$. We define the augmented parameter vector as
\begin{equation}
    \boldsymbol{\theta}=[\mathbf{u}_e^\top \quad \mathbf{p}^\top]^\top \in \mathbb{R}^{3 + 2M}
\end{equation}



Using the observations \([\mathbf{y}^{(\tau)}]_{\tau=0}^{t-1}\) and \([\textbf{y}_s^{(\tau)}]_{\tau=0}^{t-1}]\), our goal is to design the next RIS configuration \(\mathbf{k}^{(t+1)}\) while accounting for the scattering in the environment, so that the conditional \acrshort{rmse} of the estimate of \(\mathbf{u}_e\) is improved. For this purpose, the optimization task in \eqref{eq:optimization_problem} is reformulated to be processed on each instant \(t\) as follows
\begin{subequations} \label{modelingoptimization_problem}
\begin{align}
\min_{G^{(t)}(\cdot), F^{(t)}(\cdot), Q(\cdot)} & \sqrt{\mathbb{E} \left[ \|\hat{\mathbf{u}}_e^{(t)} - \mathbf{u}_e\|^2 
|[ \mathbf{y}^{(\tau)}]_{\tau=0}^{t}, [\textbf{y}_s^{(\tau)}]_{\tau=0}^{t}
\right]}, \label{modelingoptimization_objective} \\
\text{subject to}
 \label{pestimate}\quad \hat{\mathbf{p}}^{(t)} &= G^{(t)}([\textbf{y}_s^{(\tau)}]_{\tau=0}^{t}),\\
 \mathbf{k}^{(t+1)}  &= F^{(t)}\left( [\textbf{y}^{(\tau)}]_{\tau=0}^{t} \right), \label{kestimate} \\
 \hat{\mathbf{u}}^{(t)} &= Q\left( [\textbf{y}^{(\tau)}]_{\tau=0}^{t} \right), \label{uestimate}\\
 \left| \left[ \mathbf{k}^{(t+1)} \right]_n \right| &= 1, \quad \forall n \in N_{RIS}, t, \label{modulus}
\end{align}
\end{subequations}
where $\hat{\mathbf{p}}(t)$ and $\hat{\mathbf{u}}_e(t)$ denote the estimate of the \acrshort{so} locations and \acrshort{ue} location at time index $t$. At an instant $t$, let $\boldsymbol{\Gamma}^{(t)}=\mathbb{E} [ (\hat{\mathbf{u}}_e^{(t)} - \mathbf{u}_e) (\hat{\mathbf{u}}_e^{(t)} - \mathbf{u}_e)^\top
|[ \mathbf{y}^{(\tau)}]_{\tau=0}^{t}, [\textbf{y}_s^{(\tau)}]_{\tau=0}^{t}]$ be the conditional \acrshort{rmse} matrix of $\mathbf{u}_e$. 

Although the formulations in \eqref{eq:optimization_problem} and \eqref{modelingoptimization_problem} are adaptive, the objective in \eqref{modelingoptimization_problem} focuses on design in a single time frame. Specifically, the function $F^{(t)}(\cdot)$ is designed independently at each time instant to minimize the \acrshort{rmse} at the immediately following stage. This approach differs from the multi stage design in \eqref{eq:optimization_problem}, which optimizes a sequence of sensing functions ${F^{(t)}(\cdot)}_{t=0}^{T-1}$ over $T$ time frames. Computing $\Gamma^{(t)}$ is challenging as it requires high dimensional integration. Hence, we adopt the conditional \acrshort{bcrlb} as the criterion for designing the next \acrshort{ris} configuration vector, as it provides a parameter independent lower bound on the \acrshort{rmse} while incorporating posterior information from prior measurements \cite{5582316}.

At the $t$-th time frame, let the joint Bayesian \acrshort{fim} $\mathbf{J}^{(t)}$ be defined as
\begin{equation}
 J(t) =
\begin{bmatrix}
J_{uu}^{(t)} & J_{up}^{(t)} \\
J_{pu}^{(t)} & J_{pp}^{(t)}
\end{bmatrix},   
\end{equation}
with block partitions corresponding to $\mathbf{u}_e$ and $\mathbf{p}$. The entries of $\mathbf{J}^{(t)}$ are defined as
\begin{equation}
\label{fim}
\bigl[ \mathbf{J}^{(t)} \bigr]_{i,j}
= -\mathbb{E}\!\left[
\frac{\partial^2
\log f\!\left( y^{(t)}, y_s^{(t)}, \boldsymbol{\theta} \mid \{ \mathbf{y}^{(\tau)} \}_{\tau=0}^{t-1}, \{ \mathbf{y}_s^{(\tau)} \}_{\tau=0}^{t-1} \right)}
{\partial \theta_i \partial \theta_j}\right],
\end{equation}
where $f\!( y^{(t)}, y_s^{(t)}, \boldsymbol{\theta} \mid \{ y^{(\tau)} \}_{\tau=0}^{t-1}, \{ y_s^{(\tau)} \}_{\tau=0}^{t-1})$ is the conditional distribution of $y(t)$, $y_s(t)$, $\boldsymbol{\theta}$ given the available observations $\{ \mathbf{y}^{(\tau)}\}_{\tau=0}^{t-1}$ and $\{ \mathbf{y}_s^{(\tau)} \}_{\tau=0}^{t-1}$. The joint density factors according to Bayes’s rule using the posterior $f(\theta\mid.)$ and the conditional likelihoods of $\mathbf{y}^{(\tau)}$ and $\mathbf{y}_s^{(\tau)}$.
The expectation is taken with respect to aforementioned conditional distribution. The parameters $\theta_i$ and $\theta_j$ represent any two arbitrary components of the augmented parameter vector $\theta \in \mathbb{R}^{3+2M}$, encompassing both the \acrshort{ue} location and orientation and \acrshort{so} location parameters.

The conditional \acrshort{rmse} of $\mathbf{u}_e$ uses the Schur complement and is lower bounded by the \acrshort{bcrlb} as
\begin{equation}
\label{bcrlbdef}
\operatorname{tr}(\boldsymbol{\Gamma}^{(t)}) > \operatorname{tr}\Big( \big(\mathbf{J}_{uu}^{(t)} - \mathbf{J}_{up}^{(t)} (\mathbf{J}_{pp}^{(t)})^{-1} \mathbf{J}_{pu}^{(t)}\big)^{-1} \Big)
\end{equation}


The problem formulation for the \acrshort{bcrlb} for \acrshort{ue} location \acrshort{rmse} minimization with environment sensing in \acrshort{rse} is now given by
\begin{subequations}
\label{bcrlbprob}
\begin{align}
\min_{F^{(t)}(\cdot)} \quad
& \operatorname{tr}\Big( \big(\mathbf{J}_{uu}^{(t)} - \mathbf{J}_{up}^{(t)} (\mathbf{J}_{pp}^{(t)})^{-1} \mathbf{J}_{pu}^{(t)}\big)^{-1} \Big) \\
\text{subject to} \quad
& \left| \left[ \mathbf{k}^{(t+1)} \right]_n \right| = 1, \quad \forall n \in N_{RIS}, t,  \\
& \mathbf{k}^{(t+1)}  = F^{(t)}\left( [\textbf{y}^{(\tau)}]_{\tau=0}^{t}\right).
\end{align}
\end{subequations}

The problem in \eqref{modelingoptimization_problem} is addressed sequentially by repeatedly solving \eqref{bcrlbprob} to determine the \acrshort{ris} configuration across $T$ time frames. Once all $T$ stages are completed, the mapping $Q(\cdot)$ in \eqref{uestimate} is constructed using a maximum a posterior estimator to produce the final estimate of the \acrshort{ue} position $\hat{\mathbf{u}}(T)$.
\vspace{-1cm}
\subsection{BCRLB Calculation}
The joint Bayesian \acrshort{fim} in \eqref{fim} can be decomposed using the Bayes theorem as in \cite{6541985}
\begin{equation}
[\mathbf{J}^{(t)}]_{i,j} = [\mathbf{J}_D^{(t)}]_{i,j} + [\mathbf{J}_P^{(t-1)}]_{i,j},
\end{equation}
where $\mathbf{J}_D^{(t)}$ represents the incremental Bayesian \eqref{fim} derived from the latest observation, and $J_P^{(t-1)}$ corresponds to the \eqref{fim} accumulated from prior observations. 

In this scenario, the probability distribution $f ( y^{(t)} \mid \boldsymbol{\theta}, [y^{(\tau)}]_{\tau=0}^{t-1},  [\mathbf{k}^{(\tau)}]_{\tau=0}^{t} )$ and $f ( y_s^{(t)} \mid \boldsymbol{\theta}, [y_s^{(\tau)}]_{\tau=0}^{t-1},  [\mathbf{k}^{(\tau)}]_{\tau=0}^{t} )$ follow a complex Gaussian distribution, specifically $\mathcal{CN} ( (\textbf{H}_{\text{BS}-\text{U}}(\boldsymbol{\theta}))^H \mathbf{k}^{(t)}, \sigma^2 )$ and $\mathcal{CN} ( (\textbf{H}_{\text{BS}-\text{S}}(\boldsymbol{\theta}))^H \mathbf{k}^{(t)}, \sigma^2 )$. 

The entries of the incremental Bayesian \acrshort{fim} are given by
\begin{subequations}
\label{jd}
\begin{align}
[\mathbf{J}_D(t)]_{i,j} &= 
- \mathbb{E} \Bigg[ 
\frac{\partial^2\log f\Big(y^{(t)} \mid \boldsymbol{\theta}, [\mathbf{y}^{(\tau)}]_{\tau=0}^{t-1}\Big)}{\partial \theta_i \partial \theta_j}  
\Bigg] \notag \\
&\quad 
- \mathbb{E} \Bigg[ 
\frac{\partial^2\log f\Big(y_s^{(t)} \mid \boldsymbol{\theta}, [\mathbf{y}_s^{(\tau)}]_{\tau=0}^{t-1}\Big)}{\partial \theta_i \partial \theta_j}  
\Bigg] \label{first}
 \\
&= (\mathbf{k}^{(t)})^\text{H} \, \mathbb{E} \Bigg[ 
\frac{2}{\sigma^2} \,\Big( \textbf{H}_{\text{BS}-\text{U}}'(\theta_i) \,\big(\textbf{H}_{\text{BS}-\text{U}}'(\theta_j)\big)^\text{H} \notag \\
&\quad
+ \textbf{H}_{\text{BS}-\text{S}}'(\theta_i) \, \big(\textbf{H}_{\text{BS}-\text{S}}'(\theta_j)\big)^\text{H} \Big)
\Bigg]\mathbf{k}^{(t)}, 
\end{align}
\end{subequations}
where $\textbf{H}_{\text{BS}-\text{U}}'(\theta_i) = \partial \textbf{H}_{\text{BS}-\text{U}}(\theta_i)/\partial \theta_i$ and similarly for $\textbf{H}_{\text{BS}-\text{S}}'(\theta_i)$. The first expectation in \eqref{first} is taken with respect to the joint distribution $f(\boldsymbol{\theta}\mid[\mathbf{y}^{(\tau)}]_{\tau=0}^{t-1})$, $f(y^{(t)} \mid \boldsymbol{\theta}, [\mathbf{y}^{(\tau)}]_{\tau=0}^{t-1})$, $f(\boldsymbol{\theta} \mid [\mathbf{y}_s^{(\tau)}]_{\tau=0}^{t-1})$, and the second expectation is taken with respect to $f(y_s(t) \mid \boldsymbol{\theta}, [\mathbf{y}_s^{(\tau)}]_{\tau=0}^{t-1})$ assuming conditional independence of $\mathbf{y}^{(t)}$ and $\mathbf{y}_s^{(t)}$ given $\mathbf{\theta}$ and past observations.

The elements of the \acrshort{fim} derived from historical observations are expressed as
\begin{subequations}
    \label{jp}
    \begin{align}
        [\mathbf{J}_P^{(t-1)}]_{i,j} &= - \mathbb{E} \left[ \frac{\partial^2\log f \left( \boldsymbol{\theta} \mid [\mathbf{y}^{(\tau)}]_{\tau=0}^{t-1} \right)}{\partial \theta_i \partial \theta_j}  \right]\notag\\
        &\quad
        - \mathbb{E} \left[ \frac{\partial^2\log f \left( \boldsymbol{\theta} \mid [\mathbf{y}_s^{(\tau)}]_{\tau=0}^{t-1} \right)}{\partial \theta_i \partial \theta_j}  \right]\\
        &= \sum_{t' = 1}^{t-1}  - \mathbb{E} \left[ \frac{\partial^2 \log f \left( y^{(t')} \mid \boldsymbol{\theta}, [\mathbf{y}^{(\tau)}]_{\tau=0}^{t'-1} \right)}{\partial\theta_i \partial \theta_j}  \right] \notag\\ 
        &\quad
        + \sum_{t' = 1}^{t-1}  - \mathbb{E} \left[ \frac{\partial^2 \log f \left( y_s^{(t')} \mid \boldsymbol{\theta}, [\mathbf{y}_s^{(\tau)}]_{\tau=0}^{t'-1} \right)}{\partial \theta_i \partial \theta_j}  \right]\notag\\ &\quad
        -\mathbb{E} \left[ \frac{\partial^2 \log f(\boldsymbol{\theta})}{\partial \theta_i \partial \theta_j} \right]\label{second}
    \end{align}
\end{subequations}
where expectations are taken with respect to the probability distributions $ f ( y(t) \mid \boldsymbol{\theta}, [\mathbf{y}^{(\tau)}]_{\tau=0}^{t-1} )$ and $ f ( y_s(t) \mid \boldsymbol{\theta}, [\mathbf{y}_s(\tau)]_{\tau=0}^{t-1} )$. The summations over $t'$ in \eqref{second} account for the information accumulated from all past observations up to time $(t-1)$, thereby representing the total historical Fisher information contributed by prior measurements and the prior distribution of $\boldsymbol{\theta}$.

\begin{algorithm}[t]
\LinesNumbered
\caption{BCRLB Algorithm for UE localization in RSE}
\label{alg:bcrlb_ris}
\SetKwInOut{Input}{Input}
\SetKwInOut{Output}{Output}

\Input{Initial posterior distributions.}
\Output{Estimated \acrshort{ue} location and orientation $\hat{\mathbf{u}}_e^{(T)}$.}

Initialize $t = 0$\;

\While{$t <= T$}{
Generate samples from the posterior distributions
$f(\mathbf{u}_e \mid [\mathbf{y}^{(\tau)}]_{\tau=0}^{t}, [\mathbf{k}^{(\tau)}]_{\tau=0}^{t})$
and
$f(\mathbf{p} \mid [\mathbf{y}_s^{(\tau)}]_{\tau=0}^{t}, [\mathbf{k}^{(\tau)}]_{\tau=0}^{t})$\;

Compute $\mathbf{J}_D^{(t)}$ in \eqref{jd} and $\mathbf{J}_P^{(t-1)}$ in \eqref{jp}\;

Solve the optimization problem in \eqref{optproblem} to obtain the RIS configuration $\mathbf{k}^{(t+1)}$\;
}

\Return Obtain the location and orientation estimate $\hat{\mathbf{u}}_e(T)$ from the posterior distribution.
\end{algorithm}

The conditional \acrshort{fim} in \eqref{bcrlbdef} is dependent only on $\mathbf{k}^{(t)}$ through $\mathbf{J}_D(t)$, as mentioned in \cite{6541985}. The optimization problem can be given as
\begin{subequations}
\label{optproblem}
\begin{align}
&\min_{\mathbf{k}^{(t+1)}} \, \operatorname{tr}\Big( \big(\mathbf{J}_{uu}^{(t)} - \mathbf{J}_{up}^{(t)} (\mathbf{J}_{pp}^{(t)})^{-1} \mathbf{J}_{pu}^{(t)}\big)^{-1} \Big)\\
\text{subject to}
&\left| \left[ \mathbf{k}^{(t+1)} \right]_n \right| = 1, \quad \forall n \in N_{RIS}.
\end{align}
\end{subequations}

To solve this optimization problem, we employ the projected gradient descent method described in \cite{8982186}. Additionally, the posterior distribution $f \left( \boldsymbol{\theta} \mid [\mathbf{y}^{(\tau)}]_{\tau=0}^{t}, [\mathbf{k}^{(\tau)}]_{\tau=0}^{t} \right)$ is computed recursively
\begin{subequations}
\begin{align}
   f \left( \boldsymbol{\theta} \mid [y^{(\tau)}]_{\tau=1}^{t}, [y_s^{(\tau)}]_{\tau=1}^{t},[\mathbf{k}^{(\tau)}]_{\tau=1}^{t} \right) =\notag\\
   L \cdot f \left( y^{(t)} \mid \boldsymbol{\theta}, \mathbf{k}^{(t)} \right) \, f \left( \boldsymbol{\theta} \mid [y^{(\tau)}]_{\tau=0}^{t-1}, [y_s^{(\tau)}]_{\tau=1}^{t},[\mathbf{k}^{(\tau)}]_{\tau=0}^{t-1} \right), 
   \end{align}
\end{subequations}
where $L$ is the normalization constant. 

The adaptive \acrshort{bcrlb} algorithm is summarized in Algorithm \ref{alg:bcrlb_ris}. For the \acrshort{mimo} case, the conditional likelihood is vector-valued, and the Bayesian \acrshort{fim} is computed from the derivatives of the mean vectors with respect to the augmented parameter vector $\boldsymbol{\theta}$ as derived in \cite{9500437}. The incremental matrix $\mathbf{J}_D^{(t)}$ follows from the squared norm of these gradients and the Schur complement and conditional \acrshort{rmse} expressions remain unchanged. 

\section{Performance Evaluation}
\label{results}
\subsection{Simulation Setup}
The simulation parameters are summarized in Table~\ref{simulationparams}. As in Figure \ref{fig1}, in the system setup considered, the location of \acrshort{bs} and the various distributed \acrshort{ris} installations are known and fixed. The \acrshort{ue} locations \( \mathbf{u} \) are generated uniformly within the rectangular periphery in the $x$-$y$ plane with spacing $\lambda/2$. The unknown orientation $\alpha$  of the \acrshort{ue} is independently drawn from a uniform distribution over the interval $\alpha \in [0, 2\pi)$. The pilot signal \( \mathbf{x}^{(t)} \) is a deterministic scalar set to unity for all time stages. 

We consider $M$ \acrshort{so} present in the environment, each moving randomly within the rectangular region. In each simulation realization, their positions are independently drawn from a uniform distribution over the considered area. For simplicity, the positions of all \acrshort{so} are updated at the same time, while their motion remains independent. The positions of the central dipoles in each \acrshort{so} cluster are recorded in each simulation realization.  
The adjustable dipole parameters for transceivers, the environment including the \acrshort{so}, and the \acrshort{ris} such as the resonance frequency (\( f_{\text{res}} \)), the charge term (\( \chi \)) and the absorptive damping term (\( \Gamma_L \)), are set as in Table~\ref{simulationparams} \cite{9856592}. 
\begin{table}[t]
    \caption{Numerical Results Parameters}
    \centering
    \begin{tabular}{|c|c|}
        \hline
        \textbf{Label} & \textbf{Dimension} \\ 
        \hline
        \multicolumn{2}{|c|}{\textbf{Simulation Parameters}}\\
        \hline
        $M$ & {2, 4, 6}\\ 
        \hline
         $N_{\text{BS}}$& 
         4\\
        \hline
        $N_{\text{U}}$& 
         4\\
       \hline
        $N_{\text{RIS}}$& 
         {20, 60, 100}\\
        \hline
        $S_{\text{RIS}}$& 
         {2, 4, 6}\\
        \hline
        Transceivers dipole parameters & $f_{\text{res}} = 1$, $\chi = 0.5$, $\Gamma_L = 0$ \\
        \hline
        Transceivers dipole spacing &\(\lambda/2 \)\\
        \hline
        Environment and \acrshort{so} dipole parameters & $f_{\text{res}} = 10$, $\chi = 50$, $\Gamma_L = 10^4$ \\
        \hline
        Environment and \acrshort{so} dipole spacing &\(\lambda/4 \)\\
        \hline
        RIS dipole parameters & $f_{\text{res}} \in \{0.8, 5\}$, \\ & $\chi = 0.2$,$\Gamma_L = 0.03$\\
        \hline
        RIS dipole spacing &\(\lambda/4 \)\\
        \hline
        \multicolumn{2}{|c|}{\textbf{Parameters of the \acrshort{bilstm} Network}}\\
        \hline
        $B$ & [32,64,128]\\
        \hline
        $N_L$ & [2,4,6]\\ \hline
        $d_s$ & [25,35,45] \\ \hline
        $\eta$ & [0.001,0.01,0.005,0.05]\\
        \hline
         Maximum epochs & 200\\
        \hline
        $L_n$&4\\
        \hline
    \end{tabular}
    \label{simulationparams}
\end{table}
\subsection{Proposed and Baseline Approaches}
The proposed approach for \acrshort{ue} location and orientation estimation with environment sensing is  implemented with the parameters settings given in Table~\ref{simulationparams}. \acrfull{bo} \cite{hutter2019automated, NIPS2012_05311655} is used to tune the hyperparameters of the \acrshort{bilstm} networks before training by selecting the number of BiLSTM layers \( N_L \), the hidden dimension \( d_s \), the learning rate \( \eta \), and the batch size \( B \) to balance localization accuracy and computational cost \cite{11359994}. The network is trained with Adam optimizer \cite{Kingma2014AdamAM} on 2,136,000 samples over 200 epochs. The training labels for scattering estimation \acrshort{bilstm} are of dimension $2M$ and for \acrshort{ue} localization \acrshort{bilstm} are \acrshort{ue} location and orientation. After training, the model autonomously selects a sequence of \acrshort{ris} configurations, \acrshort{bs} and \acrshort{ue} beamforming vectors using sequential scattering estimates and received pilot measurements at \acrshort{bs} to accurately localize \acrshort{ue}s within the coverage area. The performance of the proposed approach is evaluated against the following benchmarks.

\textit{Beamforming free \acrshort{bilstm} network}\cite{11359994}: The adaptive \acrshort{ris} configuration is designed over $T$ time frames while accounting for scattering in the environment with the overall objective of estimating location and orientation of the \acrshort{ue}. The performance of the approach is quantified with the \acrshort{rmse} loss in location estimation as in \eqref{lossu} and scattering estimation as in \eqref{lossp}.  

\textit{\acrshort{bcrlb} based modeling and optimization}: Here an analytical model is constructed using \acrshort{bcrlb} as the optimization metric at every time frame, as discussed in Section \ref{analytical}. At each time instant $t$, the posterior distribution of the unknown \acrshort{ue} location and orientation and \acrshort{so} is refined using the pilot signals received previously, which in turn yields an updated conditional \acrshort{bcrlb}. Based on this bound, the \acrshort{ris} configuration vector $\mathbf{k}$ is optimized to reduce the conditional \acrshort{bcrlb}. To determine posterior distributions, the $35 \lambda \times  55 \lambda$ service region in the $x$–$y$ plane is discretized into $350 \times 550$ grid cells.

\textit{\acrshort{ue} localization \acrshort{bilstm} only }\cite{10373816, 10279094}: This approach relies on a \acrshort{ue} localization \acrshort{bilstm} network only for both \acrshort{ue} location and orientation estimation and \acrshort{ris} configuration, \acrshort{bs} and \acrshort{ue} beamforming vector design, while neglecting the estimation of environmental scattering in \acrshort{rse} with scattering estimation \acrshort{bilstm} network. The performance of the approach is quantified with the \acrshort{rmse} loss in location estimation as in \eqref{lossu}.

\subsection{Results}
\begin{figure}[t]
\centering
\subfloat[UE location estimation.\label{fig:loc}]{
    \includegraphics[width=\columnwidth]{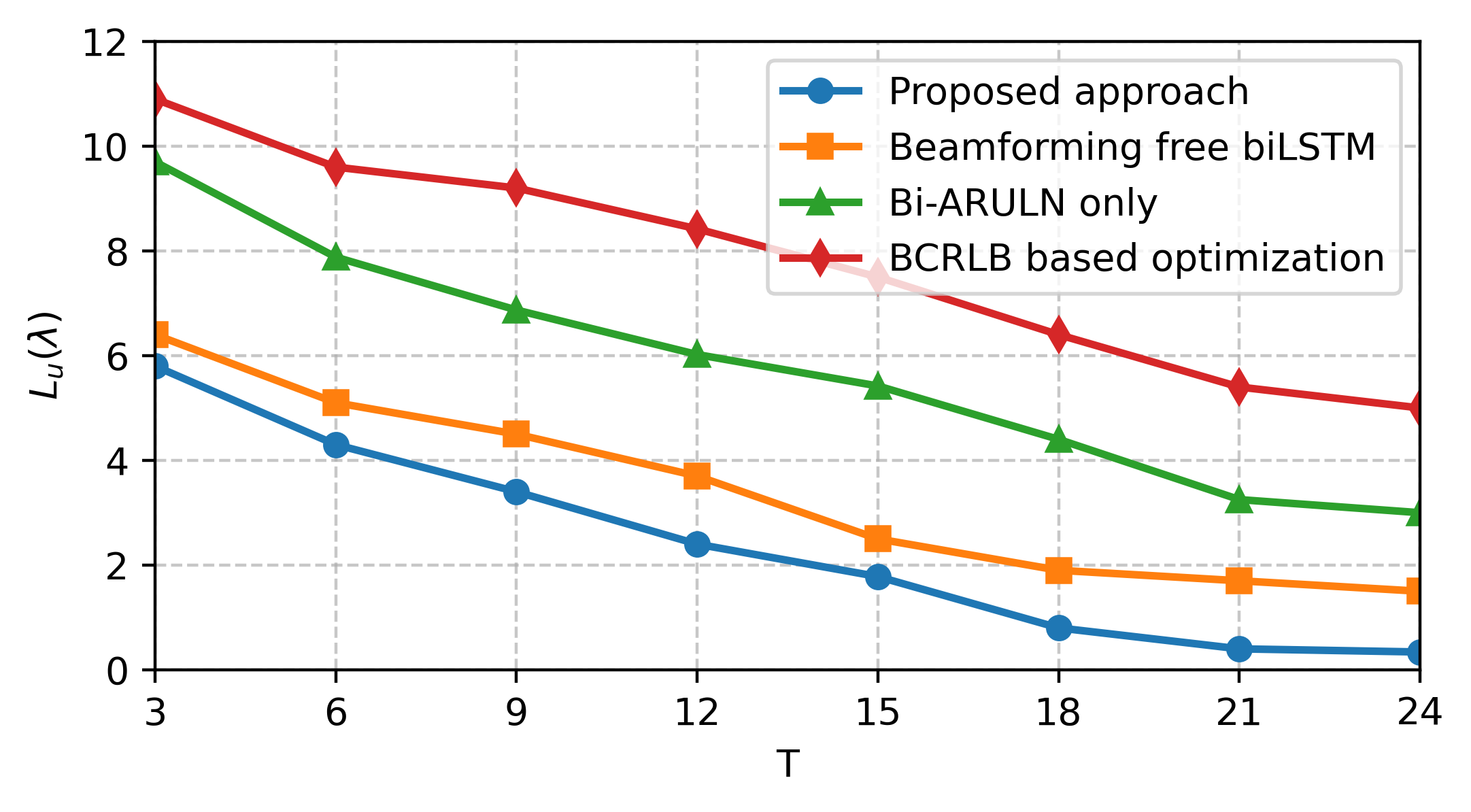}
}
\\
\subfloat[UE orientation estimation.\label{fig:orient}]{
    \includegraphics[width=\columnwidth]{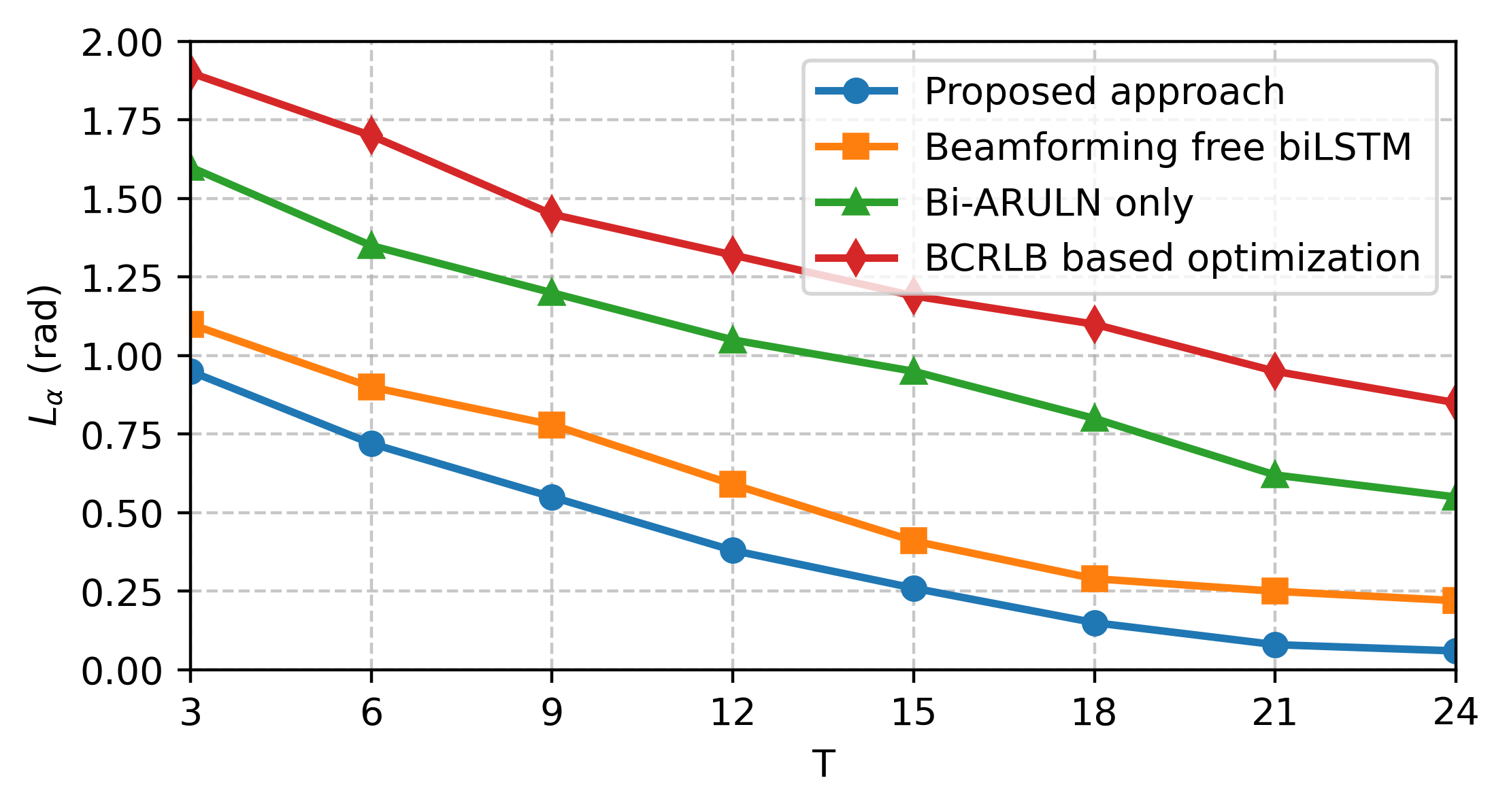}
}
\caption{RMSE of \acrshort{ue} localization versus time frames $T$ for $N_{\text{RIS}}=100$, SNR$=30$dB and $M=4$.}
\label{resultsfig1}
\end{figure}
\begin{figure}[t]
\centering
\subfloat[UE location estimation.\label{fig:snr_loc}]{
    \includegraphics[width=\columnwidth]{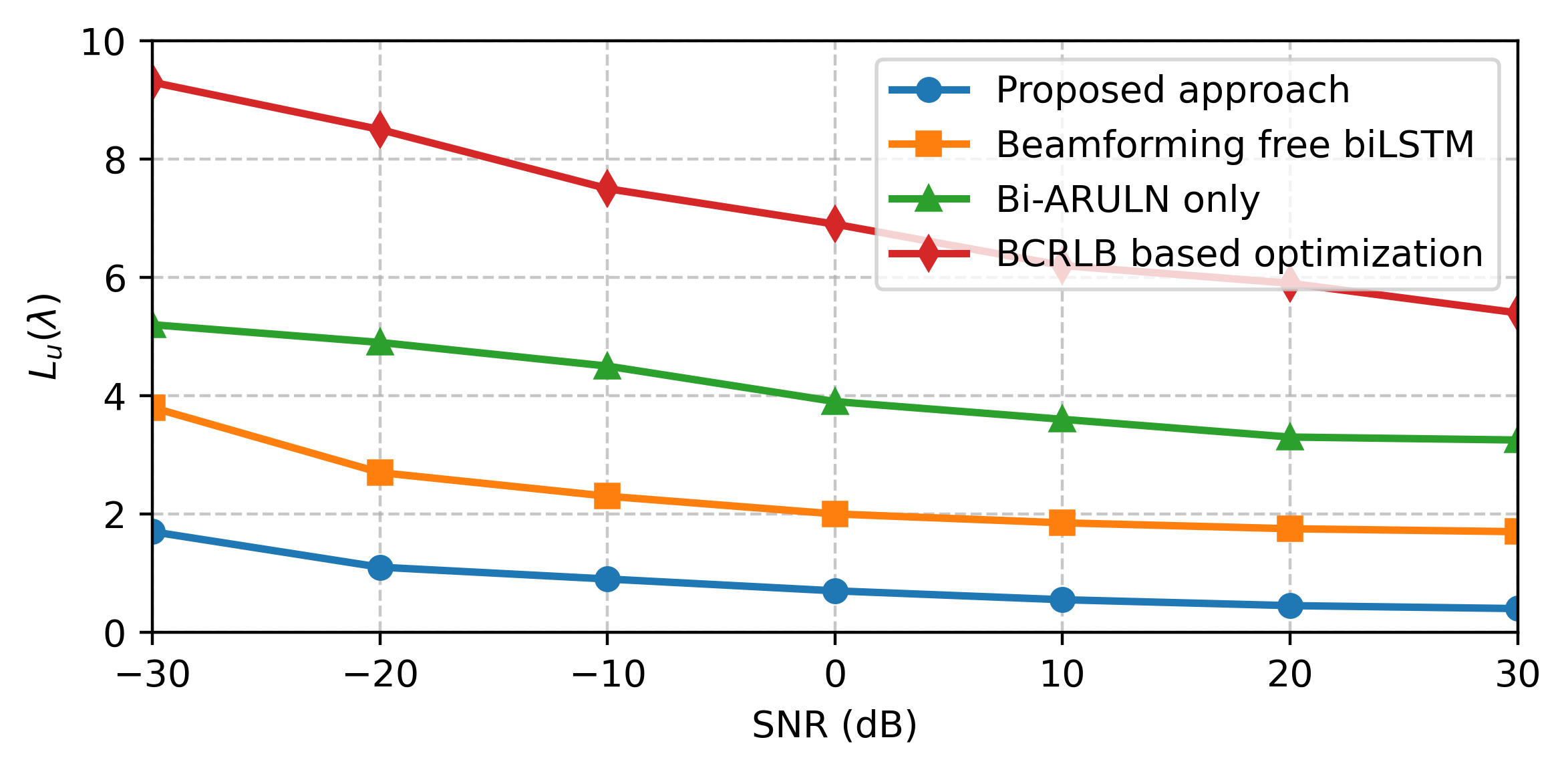}
}
\\
\subfloat[UE orientation estimation.\label{fig:snr_orient}]{
    \includegraphics[width=\columnwidth]{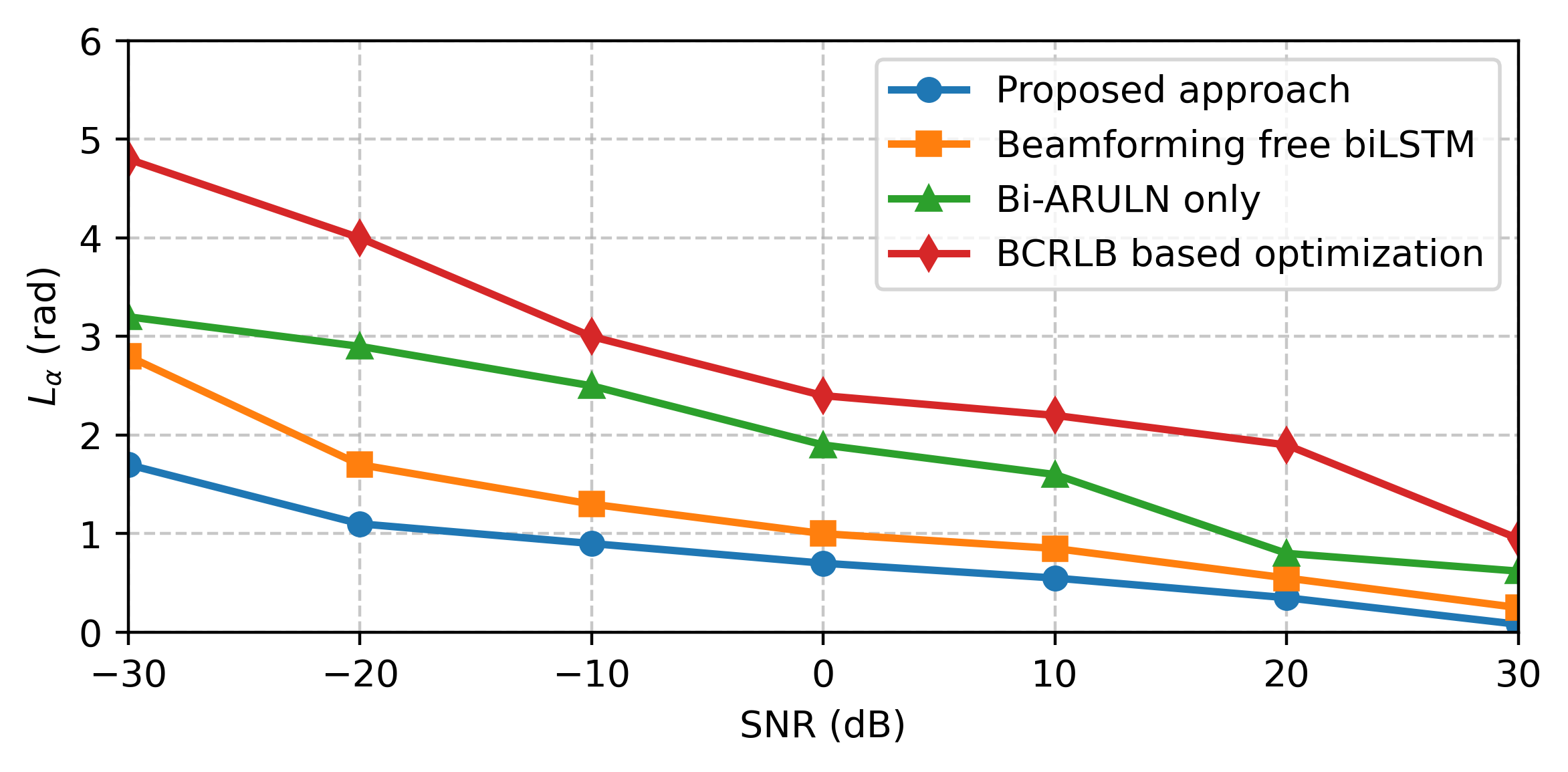}
}
\caption{RMSE of \acrshort{ue} localization versus SNR for $N_{\text{RIS}}=100$, $T=21$ and $M=4$.}
\label{resultsfig2}
\end{figure}
\begin{figure}[t]
    \centering
    \includegraphics[width=1\linewidth]{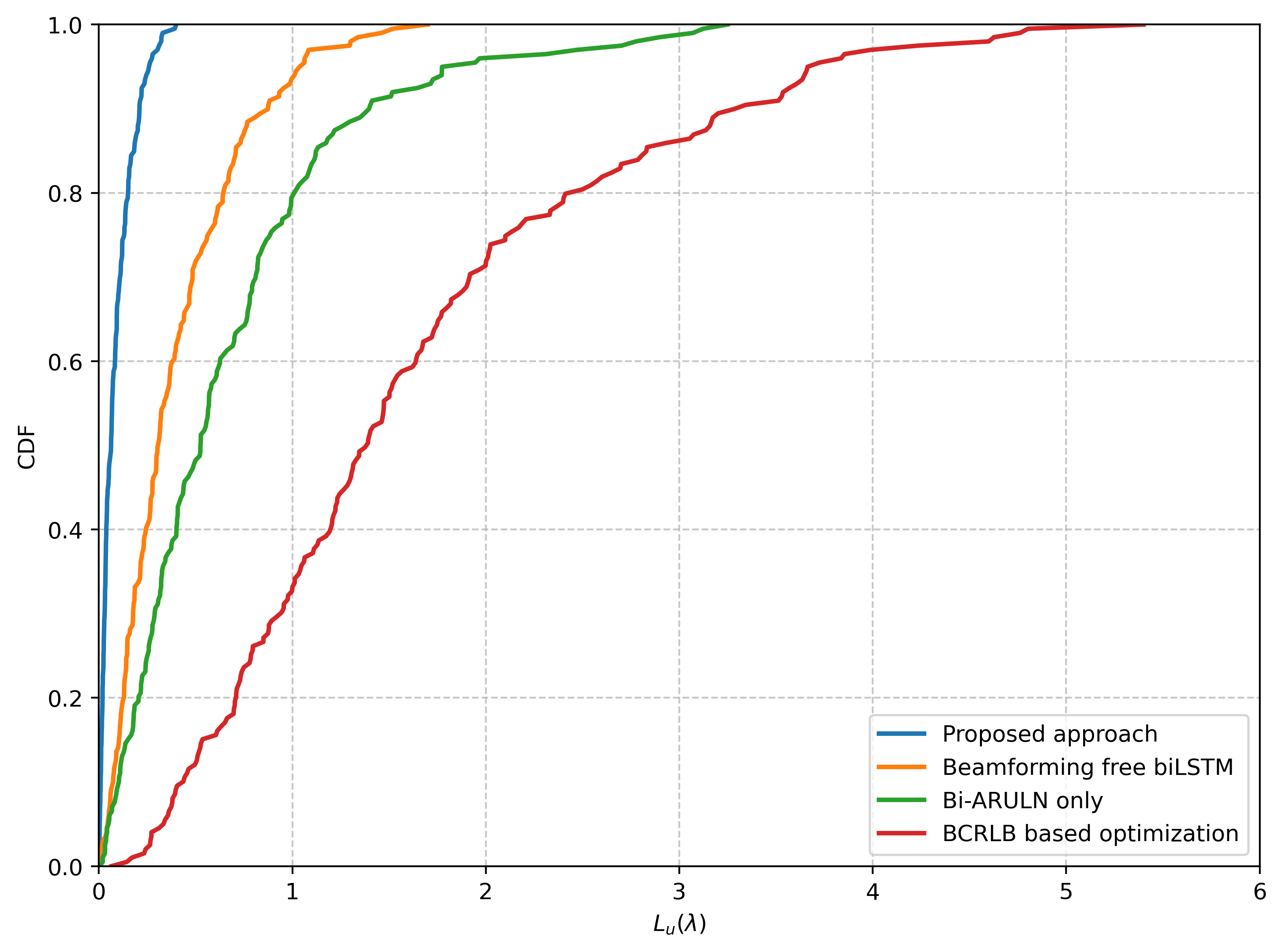}
    \caption{CDF of proposed and baseline approaches versus RMSE of \acrshort{ue} localization for $N_{\text{RIS}}=100$, $T=21$, SNR$=30$dB and , $M=4$.}
    \label{resultsfig3}
\end{figure}
\begin{figure*}[ht]
\centering
\subfloat[Left wall only.\label{onewall}]{
    \includegraphics[width=0.22\textwidth]{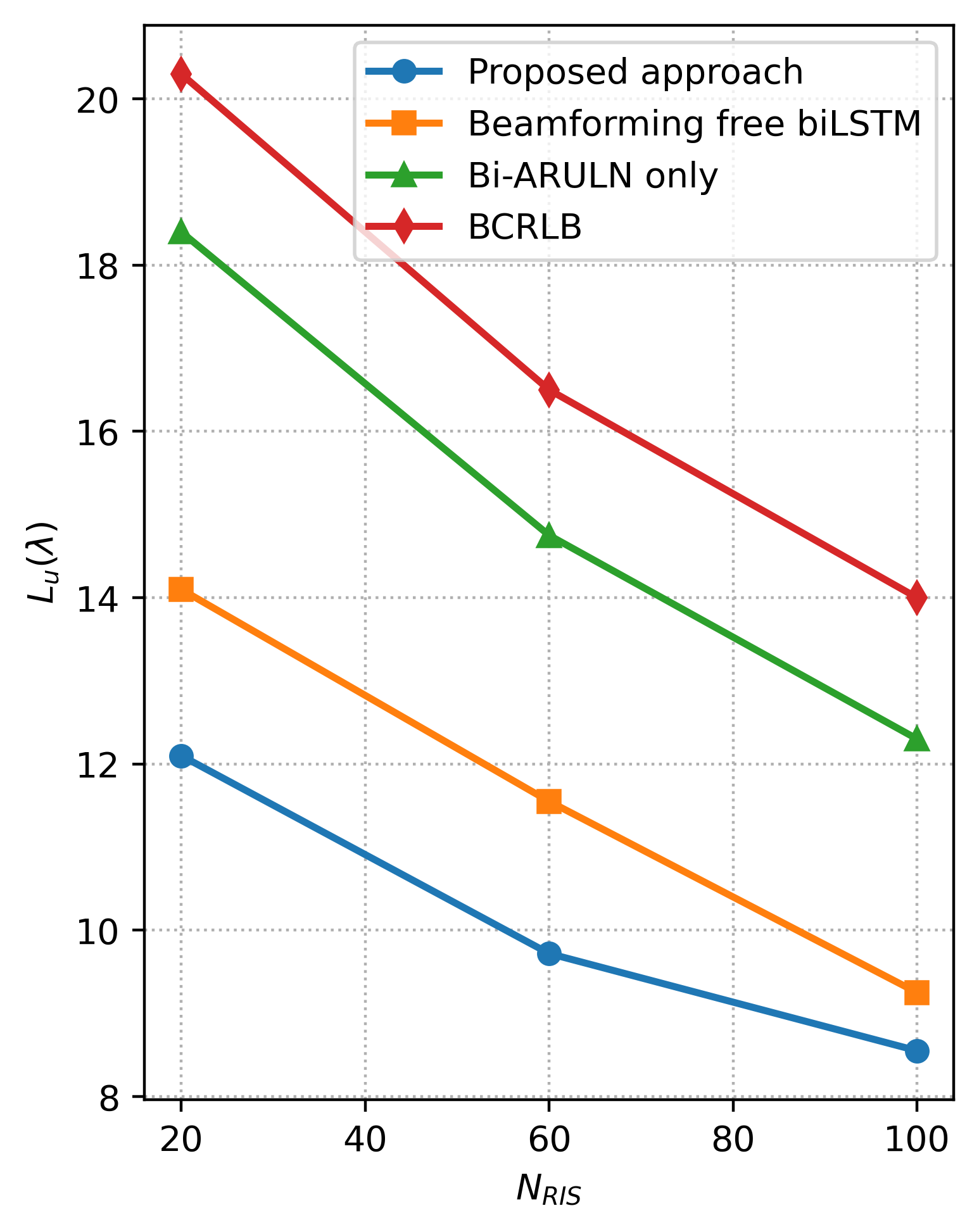}
}
\hfill
\subfloat[Left and Right wall.\label{2wall}]{
    \includegraphics[width=0.22\textwidth]{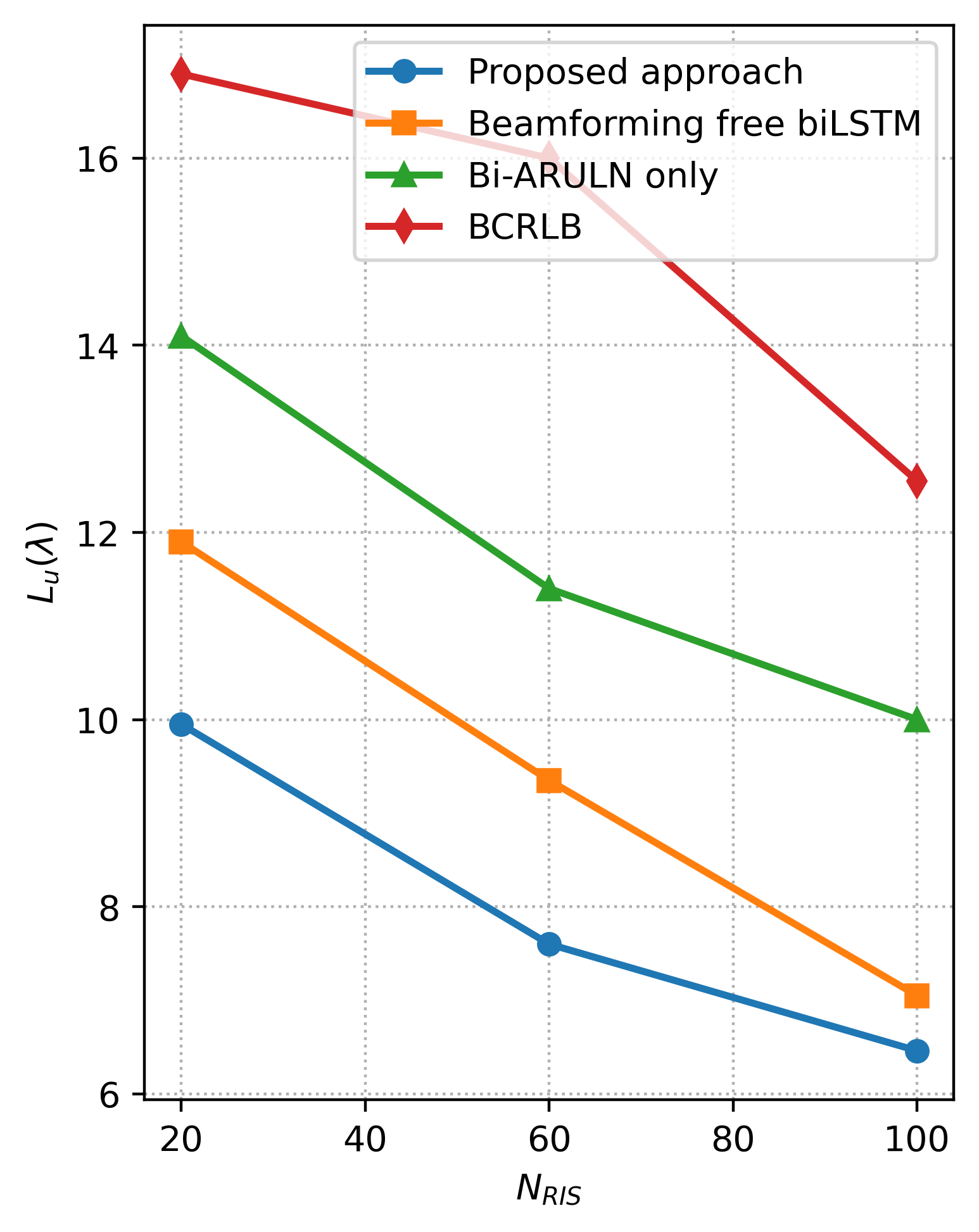}
}
\hfill
\subfloat[Left, right and top wall.\label{3wall}]{
    \includegraphics[width=0.22\textwidth]{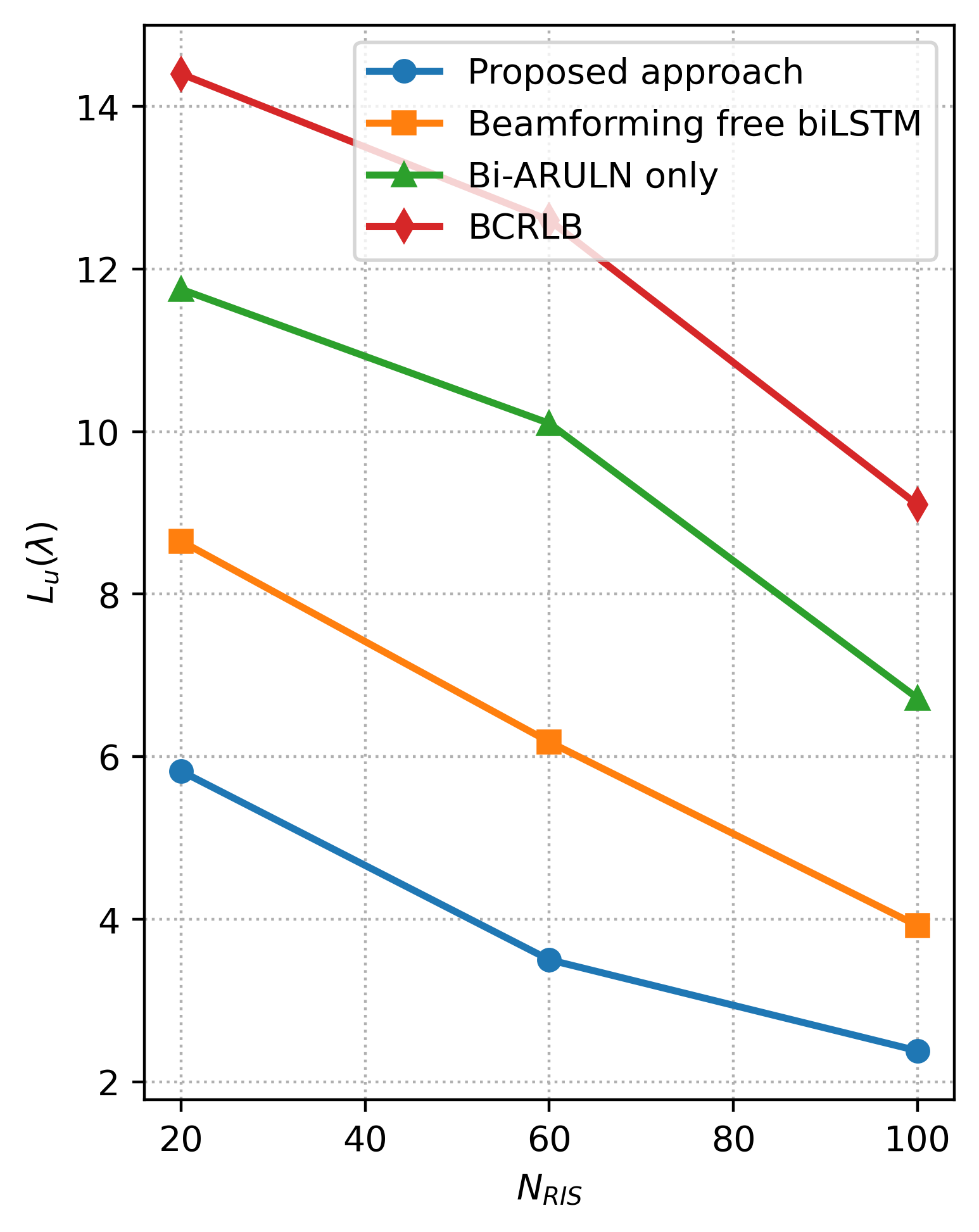}
}
\hfill
\subfloat[All 4 walls.\label{4wall}]{
    \includegraphics[width=0.22\textwidth]{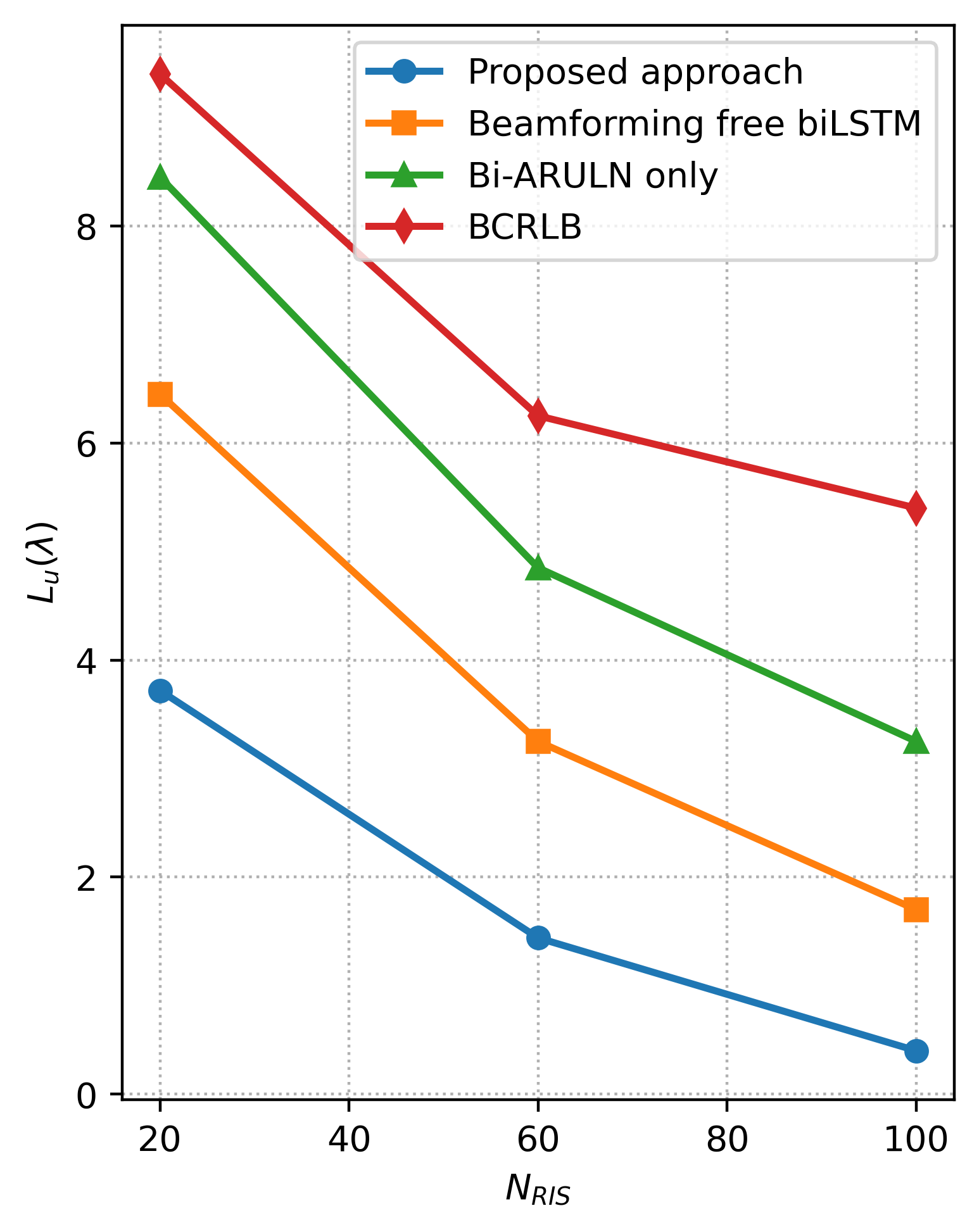}
}
\caption{RMSE of \acrshort{ue} localization versus $N_{\text{RIS}}$ under various distributed RIS installations with SNR$=30$dB, $T=21$, $M=4$.}
\label{resultsfig4}
\end{figure*}

\begin{figure}[ht]
    \centering
    \includegraphics[width=1\linewidth]{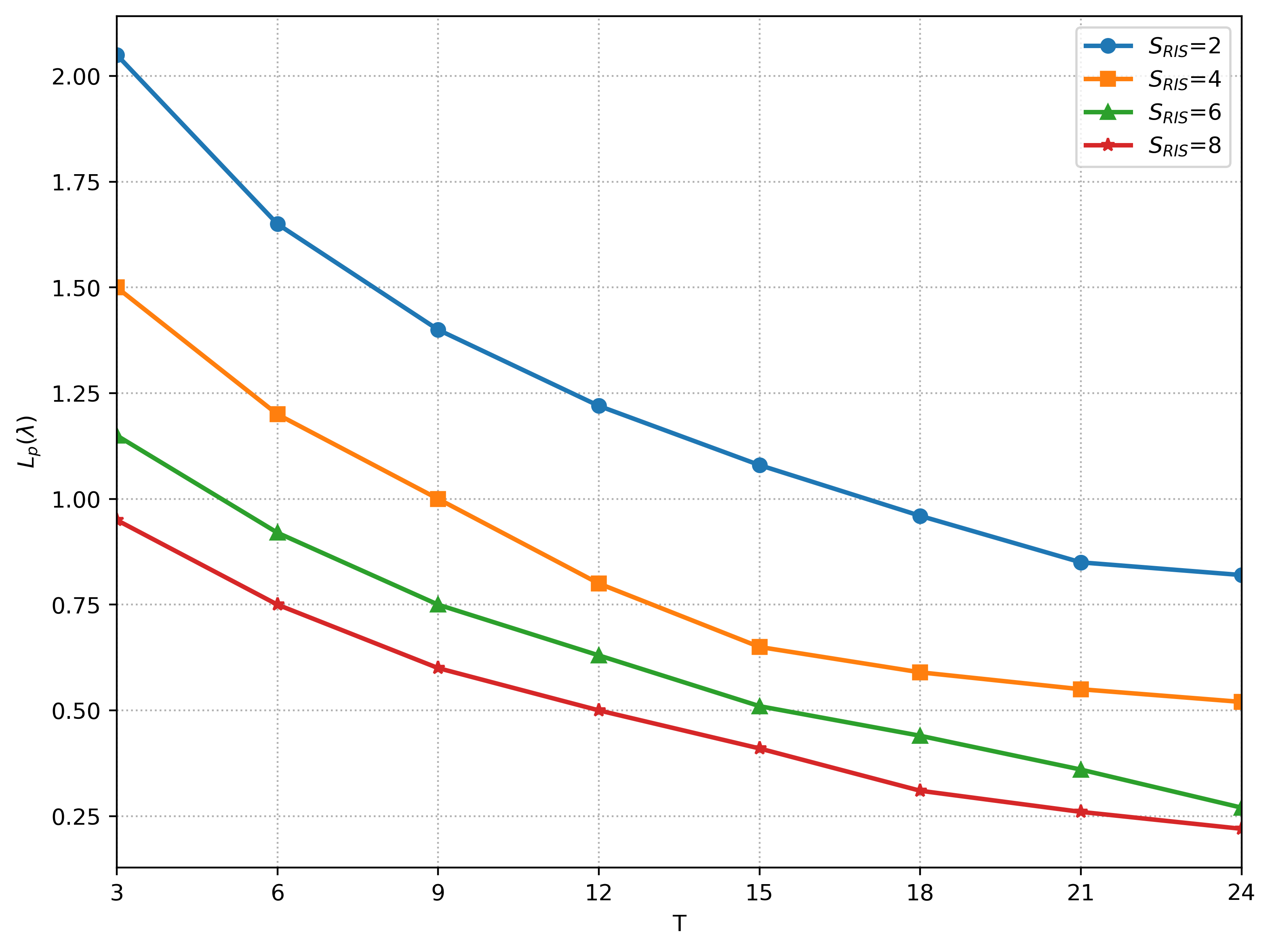}
    \caption{RMSE of \acrshort{so} localization versus time frames $T$ for different number of $S_{\text{RIS}}$ for SNR$=30$dB, $N_{\text{RIS}}=100$, $M=4$.}
    \label{resultsfig5}
\end{figure}

\begin{figure}[t]
\centering
\subfloat[UE location estimation.\label{ueloc}]{
    \includegraphics[width=\columnwidth]{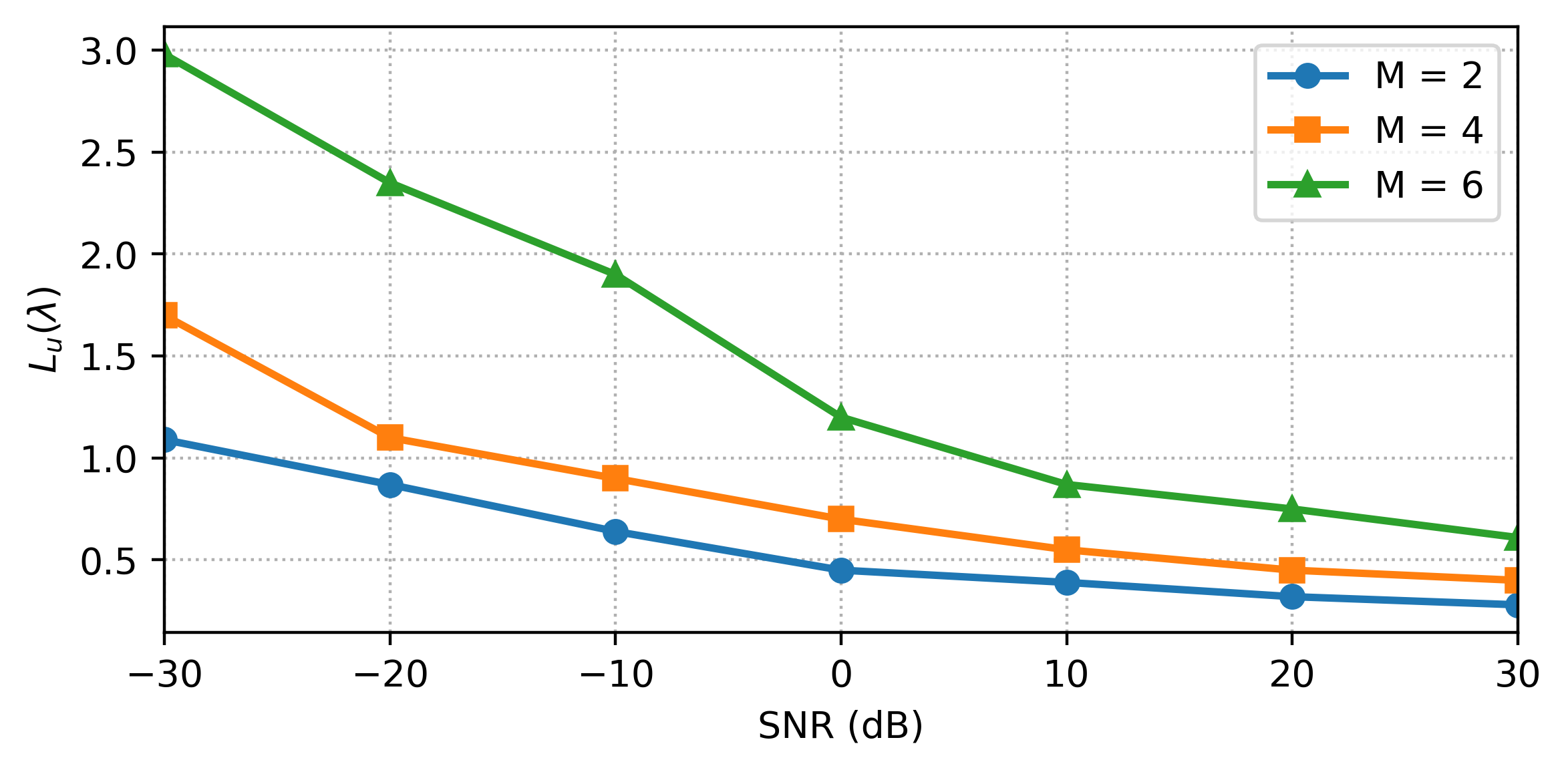}
}
\\
\subfloat[UE orientation estimation.\label{ueorien}]{
    \includegraphics[width=\columnwidth]{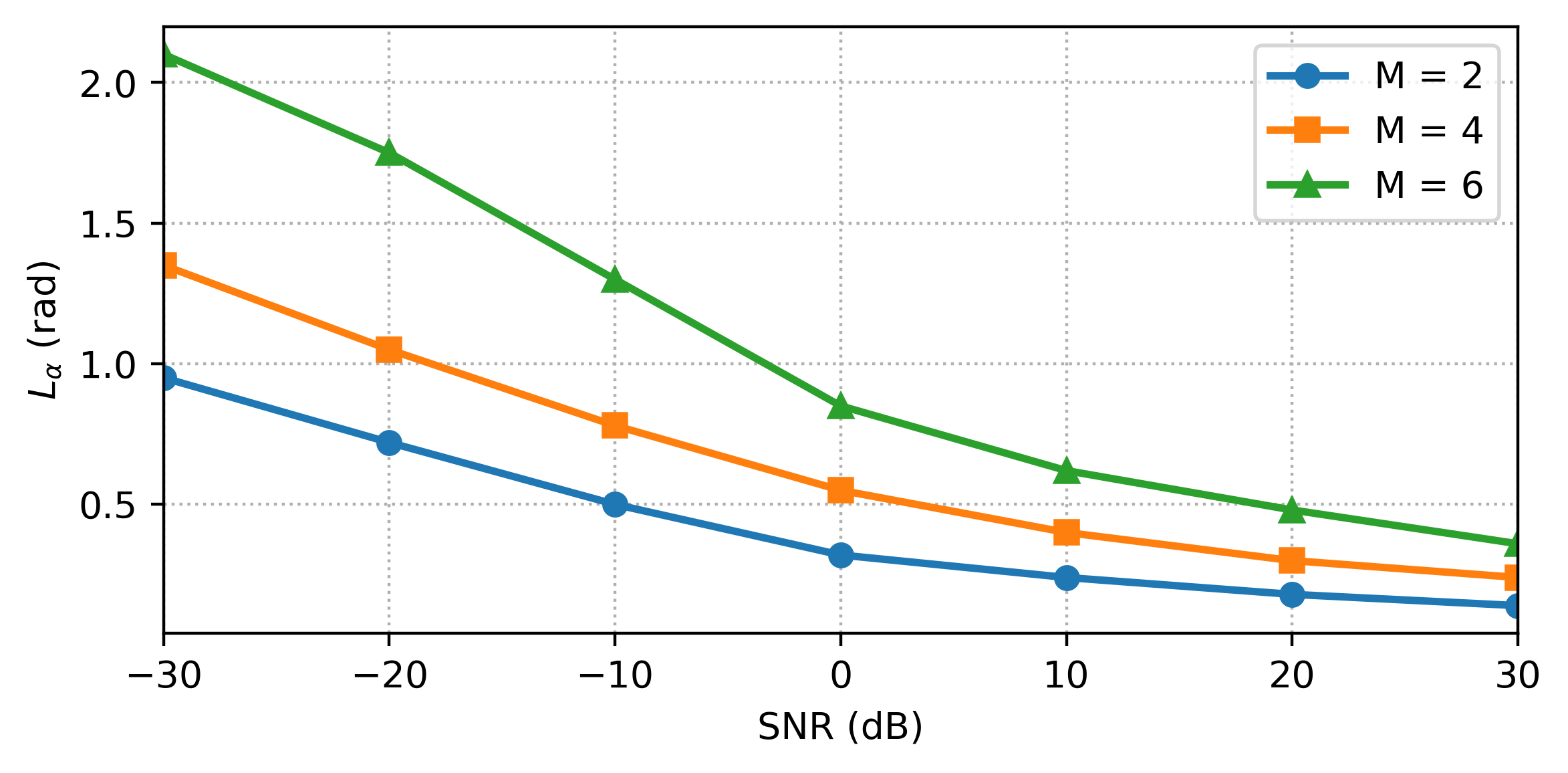}
}
\\
\subfloat[\acrshort{so} location estimation.\label{soloc}]{
    \includegraphics[width=\columnwidth]{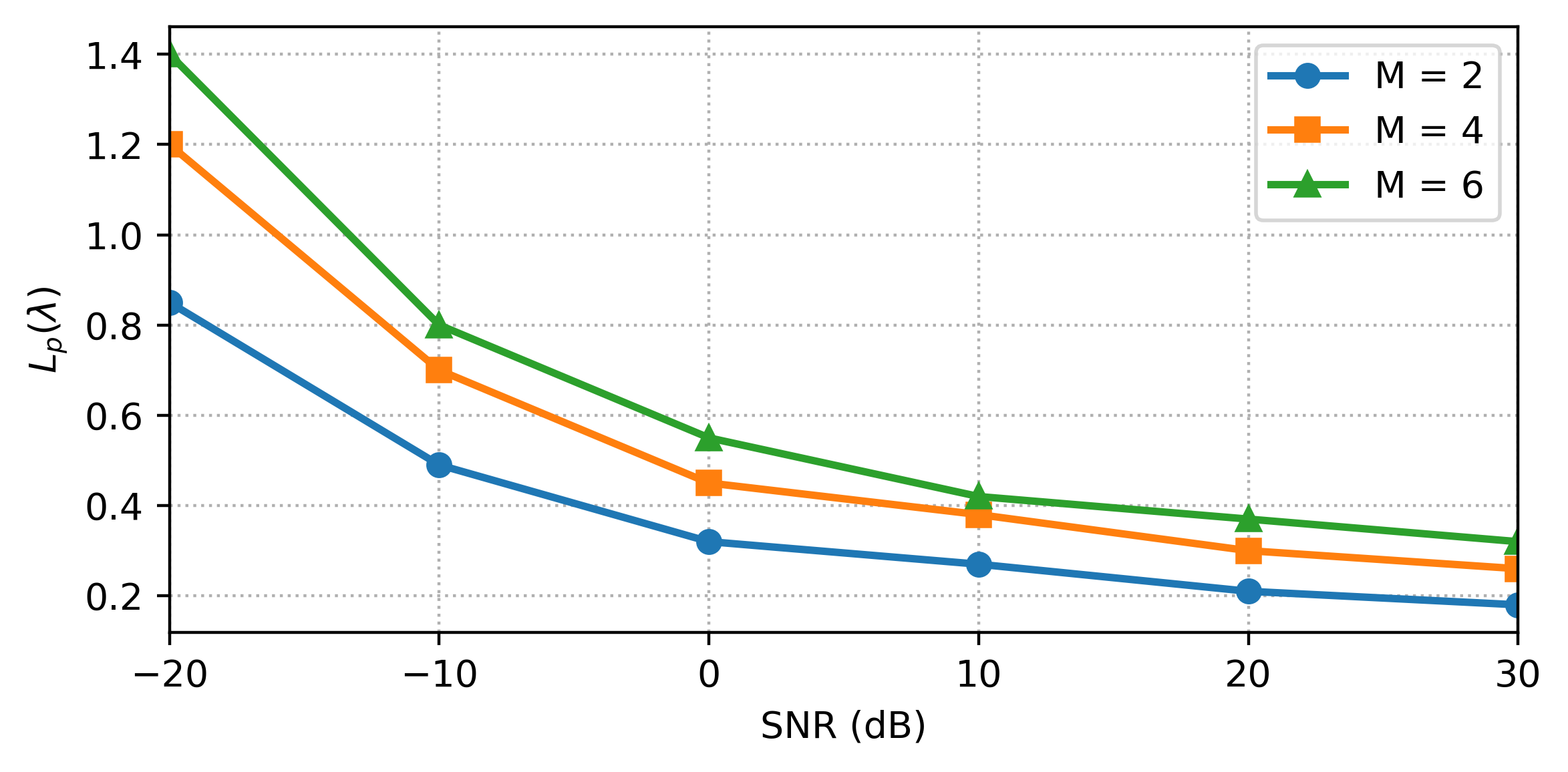}
}
\caption{RMSE of \acrshort{ue} and \acrshort{so} localization versus SNR for different number of \acrshort{so} for $N_{\text{RIS}}=100$, $T=21$, $S_{\text{RIS}}=8$.}
\label{resultsfig6}
\end{figure}
We first study the \acrshort{ue} localization performance against the number of time frames $T$ in Figure \ref{resultsfig1}. It is observed that the proposed approach has a consistently lower \acrshort{rmse} for \acrshort{ue} location and orientation estimation against all benchmark approaches for all $T$. The performance of all approaches improve with $T$ but the proposed approach outperforms both the adaptive beamforming free \acrshort{bilstm} approach and the \acrshort{ue} localization \acrshort{bilstm} only approach that does not take into account the scattering in the environment. 

It is also observed that the \acrshort{bcrlb} based adaptive \acrshort{ris} optimization approach does not perform well in comparison to the learning based approaches. It does not serve as an optimal design approach in \acrshort{rse}. This performance gap can be attributed to the difference in the design strategy, i.e., the adaptive \acrshort{ris} configuration design in our approach in \eqref{eq:optimization_problem} takes the liberty of estimating the scattering in \acrshort{rse} and designing the configuration vector across $t$ to minimize the \acrshort{rmse} of \acrshort{ue} localization at $T$, i.e., $\{G^{(t)}(\cdot), F^{(t)}(\cdot)\}_{t=0}^{T-1}$ . On the other hand. Modeling and optimization based on \acrshort{bcrlb} \eqref{modelingoptimization_problem} designs the configuration vector at each time frame $t$, i.e., $G^{(t)}(\cdot), F^{(t)}(\cdot)$. The \acrshort{bcrlb} based modeling and optimization can serve as a lower bound on \acrshort{rmse} of \acrshort{ue} localization when the number of available measurements is limited and at low \acrshort{snr} \cite{4781793}. 

We examine the \acrshort{ue} localization performance, quantized with \eqref{lossu}, against \acrshort{snr} in Figure \ref{resultsfig2}. The proposed approach exhibits the lowest \acrshort{rmse} in the range of \acrshort{snr} compared to the other learning based and analytical benchmarks. This demonstrates the significance of the proposed algorithm in utilizing the previous measurements to design \acrshort{ue}, \acrshort{bs} beamforming vectors and \acrshort{ris} configuration vector for next time frame with quantization of the scattering contribution form \acrshort{rse}. 

Next, we analyze \acrshort{cdf} of the \acrshort{ue} localization error for the proposed method and the baselines, as illustrated in Figure \ref{resultsfig3}. The proposed approach achieves the most favorable distribution, with the majority of localization errors clustered at low RMSE values, reflecting strong accuracy and robustness. The beamforming free \acrshort{bilstm} and \acrshort{ue} localization \acrshort{bilstm} only approach also delivers competitive performance, though with slightly wider error dispersion. In comparison, the \acrshort{bcrlb} based optimization performs significantly worse. Its more gradually increasing \acrshort{cdf} curve reveals larger and more inconsistent localization errors. In general, these findings demonstrate that adaptive learning driven design approaches markedly improve localization accuracy, whereas non adaptive methods or adaptive solutions that neglect scattering effects lead to increased uncertainty in dynamic \acrshort{rse} scenarios.

We study the impact of distributed \acrshort{ris} deployment and the number of \acrshort{ris} elements on \acrshort{ue} localization \acrshort{rmse} in Figure \ref{resultsfig4}. In the system model shown in Figure \ref{fig1}, we change the \acrshort{ris} deployments across walls. The single wall deployment in Figure \ref{onewall} consistently results in the highest \acrshort{rmse}, since the reflections originate from a single direction, limiting angular diversity and spatial observability despite the presence of a multipath propagation. Increasing the number of \acrshort{ris} elements enhances signal gain, but the lack of directional richness limits the achievable localization accuracy. When \acrshort{ris} panels are mounted on two walls in Figure \ref{2wall}, localization \acrshort{rmse} is significantly reduced, as reflections from multiple directions enhance the angular resolution to mitigate localization ambiguity and provides the \acrshort{bs} with more spatial information about the \acrshort{ue}. As shown in Figure \ref{3wall}, extending the deployment to three walls provides near-surround coverage, producing richer multipath propagation, controlled by \acrshort{ris}, and more pronounced RMSE reduction as \acrshort{ris} density increases.
The four wall deployment, in Figure \ref{4wall}, achieves the lowest \acrshort{rmse} overall by providing maximal spatial diversity. At large \acrshort{ris} densities, the \acrshort{rmse} gains are not very significant, indicating diminishing returns as system performance approaches the fundamental accuracy limits imposed by noise and channel estimation uncertainty in \acrshort{rse}. The proposed method consistently achieves the lowest \acrshort{rmse} across all $N_{\text{RIS}}$ and wall deployment cases, followed by the beamforming-free \acrshort{bilstm}, \acrshort{ue} localization \acrshort{bilstm} only model and the \acrshort{bcrlb} based baseline. 

The \acrshort{ris} arrays in \acrshort{rse} serve as controlled perturbers and distributing them across multiple walls with centralized control provides angular and spatial diversity, which improves the identifiability of the measurements and reduces the estimation error.
Thus, it is observed that in addition to the \acrshort{ris} configuration design, the \acrshort{ris} density and spatial distribution are critical in \acrshort{rse}, with geometry playing a greater role once sufficient elements are deployed. Together, they play a significant role in achieving the \acrshort{ue} localization accuracy objective. 

We investigate the influence of the number of sensing \acrshort{ris} elements \( S_{\text{RIS}} \) on scattering estimation performance using the scattering estimation \acrshort{bilstm}, where the estimation accuracy is quantified by the loss defined in \eqref{lossp}, as illustrated in Figure \ref{resultsfig5}. The results show a consistent reduction in \acrshort{rmse} as \( S_{\text{RIS}} \) increases across all the time frame lengths considered \( T \). This behavior highlights the critical role of the density of the sensing elements in capturing the spatial characteristics of a \acrshort{rse}. A larger number of \( S_{\text{RIS}} \) provides enhanced spatial sampling of the impinging wavefronts, which improves multipath resolvability and allows the neural network to better infer the underlying scattering. The decrease in \acrshort{rmse} with increasing \( T \) for all values of \( S_{\text{RIS}} \), indicates that longer temporal observation windows allow the scattering estimation \acrshort{bilstm} network to better exploit temporal correlations in the scattering process. However, for small \( S_{\text{RIS}} \), the reduction in RMSE with increasing \( T \) is relatively gradual, suggesting that limited spatial sampling constrains the effectiveness of temporal averaging. In contrast, when \( S_{\text{RIS}} \) is large, the combined availability of rich spatial measurements and extended temporal information results in a steeper RMSE decay and a lower overall estimation error.

These improvements in scattering estimation accuracy translate into more reliable \acrshort{ris} configuration design decisions, as the learned scattering characteristics reflect the underlying propagation environment more accurately. Consequently, enhanced scattering estimation enabled by larger \( S_{\text{RIS}} \) leads to improved \acrshort{ris} phase design and, ultimately, more precise \acrshort{ue} localization, since reliable characterization of the \acrshort{rse} allows the adaptive \acrshort{ris} control and learning based proposed approach to operate with lower ambiguity and greater robustness.

The impact of the number of \acrshort{so} \( M \) in a \acrshort{rse} on both \acrshort{ue} and \acrshort{so} localization performance is illustrated in Figure~\ref{resultsfig6}. The results indicate that increasing \( M \) leads to higher estimation \acrshort{rmse} across all considered \acrshort{snr} values in \acrshort{ue} location, orientation, and \acrshort{so} localization. This performance degradation is primarily caused by the increased complexity of the scattering environment. As the number of randomly moving \acrshort{so} increases, the dimensionality of the estimation problem increases, resulting in a larger set of unknown parameters that must be jointly estimated. In rich scattering conditions, multiple \acrshort{so} generate densely overlapping multipath components with closely spaced delays and angles. This strong multipath coupling reduces the separability of individual scattering contributions and increases inter-path interference, thereby elevating uncertainty in the learning-based scattering estimation process in the proposed approach. The resulting estimation errors propagate to subsequent \acrshort{ue} localization stages. This effect is especially pronounced at low SNR, where noise further obscures the complex multipath structure and amplifies the increase in \acrshort{rmse} with larger \( M \). 

Focusing on \acrshort{ue} localization performance in Figure~\ref{ueloc}, the \acrshort{ue} location \acrshort{rmse} is observed to increase significantly with the number of \acrshort{so} present. Although higher \acrshort{snr} improves localization accuracy for all values of \( M \), the relative performance gap between different scattering densities remains largely unchanged, indicating that increased \acrshort{snr} alone cannot fully compensate for the loss of spatial resolution caused by dense scattering. A similar trend is observed for \acrshort{ue} orientation estimation in Figure~\ref{ueorien}, where additional SOs introduce greater angular ambiguity and a stronger correlation among multipath components. The effect of \( M \) on \acrshort{so} localization is shown in Fig.~\ref{soloc}. As the number of scattering objects increases, the \acrshort{rmse} of \acrshort{so} location estimates also increases due to mutual interference among the scattering paths and the reduced identifiability of individual \acrshort{so} contributions. While increasing SNR mitigates this degradation to some extent, performance improvements gradually saturate at moderate to high SNR levels, revealing fundamental limitations imposed by dense and dynamic \acrshort{rse}.

In general, this result demonstrates that although the proposed approach remains robust in \acrshort{rse}, the number of \acrshort{so} plays a critical role in determining the achievable localization accuracy. Moderate \acrshort{so} densities provide a favorable balance between multipath diversity and estimation complexity, whereas excessively dense and randomly varying scattering introduces ambiguity that constrains further performance gains. This highlights the importance of adaptive \acrshort{ris} configuration and accurate scattering modeling for reliable \acrshort{ue} localization and \acrshort{so} sensing in dynamic RSEs.

\section{Conclusion}
\label{conclusion}
This paper presents a learning-based \acrshort{ue} location and orientation estimation approach for \acrshort{ris}-assisted communication network operating in \acrshort{rse} to sequentially design the \acrshort{ris} configuration, the \acrshort{bs} beamforming vector and the \acrshort{ue} beamforming vector for accurate \acrshort{ue} localization while estimating scattering contributed by dynamic \acrshort{so} in the environment. Using \acrshort{ris}-assisted sensing and sequential measurements history in \acrshort{bs}, our proposed approach models temporal features and adaptively designs the \acrshort{ris} configuration, \acrshort{bs} beamforming vector, and \acrshort{ue} beamforming vector to
minimize \acrshort{ue} localization error. Numerical evaluation validates that the proposed approach effectively performs beamforming and acheieves low \acrshort{ue} localization error across varying \acrshort{ris} sizes and \acrshort{ris} deployment scenarios, \acrshort{snr} conditions and varying \acrshort{rse} conditions. 

\bibliographystyle{IEEEtran} 
\bibliography{references}

\end{document}